\documentclass[a4paper,12pt]{article}
\usepackage{amssymb}
\usepackage{amsmath}
\usepackage{epsfig}
\usepackage{graphicx}
\usepackage{slashed}
\usepackage{xcolor,color}
\usepackage{ulem}
\usepackage{subfigure}
\usepackage[toc,page]{appendix}
\usepackage[makeroom]{cancel}
\usepackage{tikz}
\usetikzlibrary{shapes.geometric,positioning,decorations.pathreplacing,decorations.pathmorphing,patterns,decorations.markings}
\usetikzlibrary{arrows.meta}

\makeatletter
\renewcommand{\theequation}{\thesection.\arabic{equation}}
\@addtoreset{equation}{section}
\makeatother
\def\be{\begin{equation}}
\def\ee{\end{equation}}
\def\bea{\begin{eqnarray}}
\def\eea{\end{eqnarray}}
\def\bgn{\begin{align}}
\def\egn{\end{align}}

\def\({\left(}
\def\){\right)}
\def\<{\left<}
\def\>{\right>}

\def\({\left(}
\def\){\right)}
\def\<{\left<}
\def\>{\right>}
\def\!{\right|}
\def\|{\left|}

\def\[{\left[}
\def\]{\right]}

\def\+{\bar}

\def\ng{{\negthinspace}}

\def\eps{{\cal{\varepsilon}}}
\def\C{{\cal{C}}}

\usepackage{hyperref}
\hypersetup{colorlinks,%
  citecolor=blue,%
  linkcolor=blue,%
  pdftex}

\begin{document}

\begin{titlepage}
\vskip1cm
\begin{flushright}
% UOSTP {\tt 1512001}
\end{flushright}
\vskip0.25cm
\centerline{
\bf \large 
Quantization of Jackiw-Teitelboim gravity with a massless scalar
} 
\vskip0.8cm \centerline{ \textsc{
 Dongsu Bak,$^{ \negthinspace  a}$  Chanju Kim,$^{ \negthinspace b,c}$ Sang-Heon Yi,$^{\negthinspace d}$} }
\vspace{0.8cm} 
\centerline{\sl  a) Physics Department \& Natural Science Research Institute}
\centerline{\sl University of Seoul, Seoul 02504 \rm KOREA}
 \vskip0.2cm
\centerline{\sl b) Department of Physics, Ewha Womans University,
  Seoul 03760 \rm KOREA}
 \centerline{\sl c) Physics Department, City College of New York, CUNY,
  New York, NY 10031 USA}
   \vskip0.2cm
 \centerline{\sl d) Center for Quantum Spacetime \&  Physics Department}
  \centerline{\sl Sogang University,  Seoul 04107 \rm KOREA}
\vskip0.4cm

 \centerline{
\tt{(\small dsbak@uos.ac.kr,\,cjkim@ewha.ac.kr,\,shyi@sogang.ac.kr})
} 
  \vspace{1.2cm}
%\centerline{\today}
%\vspace{1.75cm}
\centerline{ABSTRACT} \vspace{0.65cm} 
{
\noindent 
We study canonical quantization of Jackiw-Teibelboim (JT) gravity coupled 
to a massless scalar field. We provide concrete expressions of matter 
SL(2,{\,\bf R}) charges and the boundary matter operators in terms of 
the creation and annihilation operators in the scalar field. The matter 
charges are represented in the form of an oscillator (Jordon-Schwinger) 
realization of the SL(2,{\,\bf R}) algebra. We also show how
the gauge constraints are implemented classically, by matching explicitly
classical solutions of Schwarzian dynamics with bulk solutions. 
 We introduce $n$-point transition functions defined by
 insertions of boundary
matter operators along the two-sided Lorentzian evolution, which may fully spell out the 
quantum dynamics in the presence of matter.
For the Euclidean case, we proceed with a two-sided picture of the  disk geometry and
consider the two-sided $2$-point correlation function
where initial and final states are arranged by inserting matter operators
in a specific way. For some simple initial states, we evaluate the 
correlation function perturbatively. We also discuss some basic features of the
two-sided correlation functions with additional insertions of boundary
matter operators along the two-sided evolution.

}

%\vspace{0.75cm}
%\centerline{(\today)}
\end{titlepage}
%%%%%%%%%%%%%%%%%%%%%%
%\maketitle

%%%%%%%%%%%%%%%%%

%%%%%%%%%%%%%%%%%%%%%%%%%%%%%%%%
\section{Introduction
}\label{sec1}
%%%%%%%%%%%%%%%%%
Recently there are considerable interests in  Jackiw-Teitelboim (JT) gravity in two dimensions with a negative cosmological constant \cite{Jackiw:1984je,Teitelboim:1983ux}, which may serve  as a simple model for quantum gravity (see~\cite{Mertens:2022irh,Sarosi:2017ykf,Trunin:2020vwy} for reviews). In this model, %bulk metric becomes non-dynamical while the dilaton plays the role of the bulk metric degrees of freedom in higher dimensions and the AdS cut-off trajectory fluctuation reduces to the so-called Schwarzian dynamics.  
there are no local dynamical degrees of freedom in the bulk while all the gravity dynamics are fully reflected in the boundary fluctuations of cutoff trajectories; These boundary (particle) dynamics  are well-known to be described 
by the Schwarzian theories~\cite{Maldacena:2016upp,Jensen:2016pah,Engelsoy:2016xyb}. In addition to the Euclidean path integral approach \cite{Stanford:2017thb}, this boundary picture allows
%This particular nature in the model allows 
the canonical analysis in Lorentzian %signature 
setup \cite{Harlow:2018tqv}. %, not alone Euclidean path-integral one.  
%Even with a bulk matter field, the analysis is amenable as far as the matter field does not couple the dialton directly. 
This analysis may be extended to the case of JT gravity including  matter 
as far as the matter field does not  couple directly to the dilaton field \cite{Penington:2023dql}.

The relevant AdS$_2$ geometry %JT gravity 
in Lorentzian signature %we shall mainly interested in 
is intrinsically  two-sided involving 
%involve 
 left and right cutoff trajectories near the 
AdS$_2$ boundary (see Figure \ref{fig1}). In the context of AdS/CFT
correspondence, it is rather natural to expect that the dual boundary theory has a description based on a tensor product structure of left and right theories. Hence it appears that one is able to construct a one-sided Hilbert space out of JT gravity in a natural manner. On the contrary, it has been argued that the quantized version of JT theory  allows only  a two-sided Hilbert space 
${\cal H}$ whereas a one-sided Hilbert space cannot be defined \cite{Harlow:2018tqv,Jafferis:2019wkd},   which is coined as the {\it factorization problem} in JT gravity~\cite{Harlow:2018tqv}. There are also related issues  in higher dimensions  on how to understand the behind-horizon interactions  %the boundary dual tensor product CFT,  
from the viewpoint of $\mbox{CFT}_l \otimes \mbox{CFT}_{r}$, which
 was emphasized in~\cite{Marolf:2012xe}. 
 
%One of the interesting aspects in JT gravity with negative cosmological constant is that its boundary is two-sided just like an eternal black hole in higher dimensions. In the context of the AdS/CFT correspondence, the dual CFT on the two-sided boundary has a tensor product structure, CFT$_{L}$$\otimes$CFT$_{R}$, allowing a one-sided CFT and its Hilbert space. Hence,  at first glance, one may think that an observer living on one side of the system, who is causally disconnected from the other side of the system, can construct a one-sided Hilbert space in JT gravity. On the contrary, it has been argued that the two-sided Hilbert space, ${\cal H}$, is the appropriate one to describe the quantized version of JT gravity while a one-sided Hilbert space cannot be defined~\cite{Harlow:2018tqv,Jafferis:2019wkd}, which is coined as the {\it factorization problem} in JT gravity~\cite{Harlow:2018tqv}.  There are also related issues  in higher dimensional two-sided (eternal) black holes on how to understand the behind-horizon interactions from the boundary dual tensor product CFT,  as was emphasized in~\cite{Marolf:2012xe}. 

A recent %important 
observation says that the algebra of (bulk) operators ${\cal A}$  acting on a  two-sided Hilbert space ${\cal H}$ can be constructed in such a way that two one-sided algebras ${\cal A}_{l}$ 
and  ${\cal A}_{r}$  
are well 
%meaningfully 
defined %meaningful 
and  commute with each other %side algebra, 
preserving the causality at the level of operator algebra \cite{Leutheusser:2021qhd,Leutheusser:2021frk,Witten:2021unn,Chandrasekaran:2022cip,Chandrasekaran:2022eqq} (see also~\cite{Kolchmeyer:2023gwa}, which appeared near %we have encountered after 
completion of our work). It is noticeable that, even at the level of the algebra, %level, 
the algebra 
${\cal A}$ is not a tensor product of %two algebras 
${\cal A}_{l}$ and ${\cal A}_{r}$. 
%One of the important results in this algebraic approach, which is based on the classification of von Neumann algebra, is  that the entanglement entropy can be understood from Lorentzian signature, which was obtained from the Euclidean path-integral method.  
In the %context of 
JT gravity with %coupled to bulk 
matter, it is shown that the type of von Neumann algebra is type II$_{\infty}$ and the corresponding algebra ${\cal A}_{l/r}$ is fully %completely 
specified by the %cut-off trajectory Schwarzian 
boundary Hamiltonian %operator, 
$H_{l/r}$ and the boundary matter operator $\hat{\varphi}_{l/r}$ derived %induced 
from the bulk matter operator 
%through the bulk-boundary map in the AdS/CFT correspondence
\cite{Penington:2023dql}. 

The Schwarzian theories involve higher derivative terms. However there is no inconsistency 
since these higher derivative terms are constrained by a gauge symmetry of $\widetilde{\mbox{SL}}$$ (2,{\bf R})$ which leaves the full geometry including the cutoff boundaries  invariant. By imposing the corresponding gauge constraints with an appropriate gauge-fixing, the quantization of  JT theory with and without matter has been carried out in
 \cite{Penington:2023dql}, which will be reviewed in the following. Upon quantization,  
the reduced Hilbert space exhibits a genuinely two-sided nature while the Hamiltonian  $H_{l/r}$  generates the left and right time evolution respectively. From the viewpoint of this two-sided Hilbert space,
the full dynamics of the system may be described by the  total Hamiltonian $H_{r}+H_{l}$
with a single time parameter $u$ evolving the left and right at the same time.
In the pure JT gravity,  $H_{r}-H_{l}$ vanishes identically and merely induces a pure gauge transformation. In the presence of matter, $H_{r}-H_{l}$ becomes nontrivial and generates
a relative (boostlike) time evolution.

%One interesting feature of the reduction from JT gravity to Schwarzian dynamics is the appearance of  higher derivatives terms from two derivative Lagrangian. Of course, there is no inconsistency and the higher derivatives terms are constrained by a gauge group $\widetilde{\mbox{SL}}$$ (2,{\bf R})$ which could be thought as a remnant of bulk diffeomorphism. Because of the higher derivative nature of Schwarzian dynamics, the cut-off trajectory Schwarzian Hamiltonians, $H_{r/l}$ should respect the appropriate gauge constraints. After the gauge constraints are imposed,  $H_{l/r}$ as an element in ${\cal A}_{l/r}$,  is well-defined  as the generator for each $l/r$ side time evolution, while the well-defined Hilbert space should be two-sided one. From the viewpoint of the two-sided Hilbert space, the whole dynamics should be described by the two-sided total Hamiltonian $H_{r}+H_{l}$, while there would be another time evolution dictated by $H_{r}-H_{l}$. In pure JT gravity,  $H_{r}-H_{l}$ vanishes because of the gauge constraint so it does not produce any meaningful time evolution. However, in JT gravity coupled to bulk matter, $H_{r}-H_{l}$ produces a non-trivial time evolution which is related to the existence of the boundary matter operators $\hat{\varphi}_{r/l}$ in Schwarzian dynamics. 

For  JT theory with a massless scalar field specifically, its full general solutions  are 
presented explicitly in \cite{Bak:2021qbo}, which show some nontrivial aspects of two-sided
black holes involving the matter field. For instance, the left and right temperatures of  two-sided
black holes become different from each other. Indeed, JT theory with matter  seems to exhibit
many more nontrivial features. %unlike the case of the pure JT theory. 
In this note, we consider the explicit canonical quantization of JT gravity coupled to a massless scalar field.  Based on explicit bulk expressions, we shall provide the concrete expressions of  matter ${\mbox{SL}} (2,{\bf R})$ charges and the boundary matter operators  $\hat{\varphi}_{l/r}$.  
We also provide a comparison of  the classical and quantum dynamics with some comments on the classical realization of the gauge constraints. % in Schwarzian dynamics. 

%As can read from the classical two-sided black hole solutions to the bulk equations of motion with generic bulk matter, it is manifest that the left and right boundary states corresponding to the bulk solutions could have different temperatures. This indicates that the partition function for the two-sided Hilbert space may have two temperature parameters $\beta_{l}$ and $\beta_{r}$, not a single one $\beta$. 
%This aspect of JT gravity coupled to bulk matter, as well as  boundary algebras, is different from pure JT gravity.
%Hence, it would be interesting to explore some details of JT gravity coupled to bulk matter. In this paper we consider the explicit canonical quantization of JT gravity coupled to bulk massless scalar field. Using the explicit bulk expressions, we provide the concrete expressions of  matter ${\mbox{SL}} (2,{\bf R})$ charges and the boundary matter operators  $\hat{\varphi}_{l/r}$.  We also provide the classical solutions to bulk equations of motion and the Schwarzian dynamics with some comments on the classical aspects of the gauge constraints in Schwarzian dynamics. 

In \cite{Lin:2022zxd}, 
it was shown that
the disk partition function of  pure JT gravity \cite{Stanford:2017thb} can be reproduced from 
an evaluation based on 
the two-sided picture (see the left diagram of Figure \ref{Fig2});  There one starts the two-sided evolution  from an initial geodesic curve connecting two slightly separated boundary points in the bottom region of the disk and ends up with  a final geodesic curve between two slightly separated boundary points in the top region.  We generalize the computation to the case of JT theory with the massless  scalar field, in which one may additionally arrange  initial and final states including the matter part by 
inserting matter operators before and after the initial and final regularized curves (see the right diagram of Figure \ref{Fig2}).
% such that  appropriate initial and final matter states are arranged. 
In this note,
 we %propose the way to 
%intend to 
specify  prescribed states\footnote{The Hartle-Hawking construction on a half disk~\cite{Kolchmeyer:2023gwa} may be a good alternative for preparing the initial or final state. However, it is not clear  to us how to land there starting from our Lorentzian two-sided picture.} at the initial and final regularized 
curves generalizing the proposal in \cite{Lin:2022zxd}. 
%But %our investigation of the  two-sided correlation function in this direction does not seem
% to be satisfactory. 
%our investigation of the two-sided correlation functions in this  direction has not yielded satisfactory results.
%%{\color{blue}      The preparation of
% initial (or final) state  based on the Hartle-Hawking construction on a half disk may be a good alternative \cite{Kolchmeyer:2023gwa} but how to land there starting from our two-sided picture is not clear to us.}
One may additionally  insert boundary matter operators along the two-sided evolution, which  leads to higher two-sided   correlation functions. We  investigate %some 
basic properties of these two-sided correlation functions. % in this note.

This paper is organized as follows. In section~\ref{sec2}, we give our basic setup of JT gravity. %in our notation. 
 In section~\ref{sec3},  we review the canonical quantization of JT gravity with matter following \cite{Penington:2023dql}. In section~\ref{sec4} , we specialize in  JT gravity with a massless scalar field and  provide full details of quantization. Especially we quantize 
the bulk scalar field   %quantization of massless scalar field and
 leading to explicit  expressions of  the matter ${\mbox{SL}} (2,{\bf R})$ charges. 
%We identify the matter charges, 
We find that the matter charges $J^{m}_{i}$ form the oscillator (Jordan-Schwinger) realization of ${\mbox{SL}} (2,{\bf R})$. In the case of the massless  scalar especially, 
the mapping is given in terms of the matrices of $D^{-}_{j=1}$ representation. In section~\ref{sec5}, we consider the classical solutions of Schwarzian dynamics and its relation to bulk solutions. We also check the gauge constraints in the classical setup.  In section~\ref{sec6}, we consider  two-sided correlation functions in the presence of the matter field.
%, which will be generalized  to  the case of  two-sided correlation functions. 
We present  some explicit evaluations of the two-sided correlation functions. 
%and
% investigate 
%their elementary properties. % of the two-sided correlation functions.
%on two-sided Hilbert space when a quantum bulk matter configuration is given specifically by the number operator $N_m$. 
In the final section, we summarize our results and give some comments on future directions. 
%in the final section. 

%%%%%%%%%%%%%%%%%
\section{Jackiw-Teitelboim gravity with matter
}\label{sec2}
%%%%%%%%%%%%%%%%%
We shall consider JT gravity~\cite{Jackiw:1984je,Teitelboim:1983ux,Almheiri:2014cka} coupled to a %massless 
matter field which is described by action\footnote{Here, we have omitted a topological term which is irrelevant in our discussion below. We also set $8\pi G=1$ and the AdS radius $\ell=1$.} 
\bea
I=\frac{1}{2}\int_M d^2 x \sqrt{-g}\, \phi \left( R+2\right) + I_{surf}  + I_m(g, \varphi) \,,
\label{euclidaction}
\eea
where $\phi$ is for a dilaton field,  $\varphi$ for  the matter field,  and 
\begin{align}    \label{}
I_{surf} &=   \int_{\partial M} du \sqrt{-\gamma_{uu}}\, \phi  \, (K-1)  \,, \nonumber \\
I_m\  \,&= -\frac{1}{2}\int_M d^2 x \sqrt{-g}\,  \left( g^{ab} \nabla_a \varphi %\cdot 
\nabla_b \varphi +m^2 \varphi^2 \right)\,.
\end{align}
Here, $u$ is our boundary time coordinate and  $\gamma_{uu}$ and $K$  respectively
denote the induced metric and  the extrinsic curvature on the boundary 
$\partial M$. The equation of motion following from the dilaton % $\phi$ 
variation 
 is given by 
\begin{equation}
R+2=0\,,
\end{equation}
which fixes the metric to be AdS$_2$. % whose  radius $\ell$  is set to be unity. %$\ell=1$.  
%The other equations of motion
%are obtained from the variation of the metric $g$ and the scalar field $\varphi$, 
%respectively, as
The remaining equations of motion read
\begin{align}    \label{phieq}
\nabla_a \nabla_b \phi -g_{ab} \nabla^2 \phi + g_{ab} \phi &= -  T_{ab}\,, \\
\nabla^2 \varphi -m^2 \varphi &=0\,,  \label{massless}
\end{align}
where
$T_{ab}$ is the stress tensor of the matter field,
\begin{equation} \label{}
T_{ab} = \nabla_a \varphi \nabla_b \varphi  -\frac{1}{2} g_{ab}\left( g^{cd}\nabla_c \varphi  %\cdot 
\nabla_d \varphi +m^2 \varphi^2 \right) \,.
\end{equation}
In the global coordinates, the metric of the AdS$_2$ space is written as
\begin{equation} \label{}
ds^2 =\frac{-d\tau^2 + d\mu^2 }{\cos^2 \mu} \,,
\end{equation}
where $\mu \in [-\frac{\pi}{2},\frac{\pi}{2}]$ which is strip-shaped as depicted on the left  of
Figure \ref{fig1}.  
 A vacuum solution of the dilaton field (in a gauge-fixed form)   is given by
\begin{equation}
\phi=  \bar\phi \, L\,\,\frac{ \cos \tau  }{ \cos \mu} \,,
\label{dilatonvac}
\end{equation}
which is describing a  two-sided black hole spacetime that is left-right symmetric.

\def\xxx{

\begin{figure}[htbp]  
\vskip0.3cm 
\begin{center}
\begin{tikzpicture}[scale=1.1]
\draw[thick,blue %,-<
](-2,0)--++(0,2);
\draw[thick,blue] (-2,2)%node[left]{\large$t_l$}
--++(0,2);
\draw[thick,red %, ->
](2.15,0)--++(0,2);\draw[thick,red] (2.15,2)%node[right]%{\large$t_r$}
--++(0,2);
%\draw[blue,decoration={coil,segment length=0.5mm,amplitude=0.15mm},decorate] (-2,2) arc (180:270:2);\draw[red,decoration={coil,segment length=0.5mm,amplitude=0.15mm},decorate] (2,2) arc (0:-90:2); \draw[fill=black] (0,0) circle (0.05cm);
%\draw[decoration={zigzag},decorate](-2,0)--(2,0); 
% \draw[thick,dotted](-2,2)--++(4,0);
%\draw[decoration={zigzag},decorate](-2,4)--++(4,0); 
\draw[%violet,
decoration={zigzag,segment length=1mm,amplitude=0.5mm},decorate
](-2,4)--++(0.2,+0.35); 
\draw[decoration={zigzag,segment length=1mm,amplitude=0.5mm},decorate
](2.15,4)--++(-0.2,+0.35); 
\draw[decoration={zigzag,segment length=1mm,amplitude=0.5mm},decorate
](-2,0)--++(0.2,-0.35); 
\draw[decoration={zigzag,segment length=1mm,amplitude=0.5mm},decorate
](2.15,0)--++(-0.2,-0.35); 
%\draw[thick,brown,dotted] (-2,0)--++(4,4);
%\draw[thick,brown,dotted] (+2,0)--++(-4,4);
\draw (-2,0)--++(2,2)--++(-2,2);
\draw[dotted] (0,2)--++(2.08,2.08);\draw[dotted] (0,2)--++(2.10,-2.10);
\draw[dotted] (0.15,2)--++(-2.10,2.10);\draw[dotted] (0.15,2)--++(-2.08,-2.08);
\draw (+2.15,0)--++(-2,2)--++(2,2);
%\draw (+2,0)--++(-2,2)--++(2,2);
%\draw[blue,decoration={coil,segment length=0.5mm,amplitude=0.15mm},decorate] (6,4) arc (90:270:2);%\draw[red,decoration={coil,segment length=0.5mm,amplitude=0.15mm},decorate] (6,4) arc (90:-90:2); %\draw[fill=black] (6,0) circle (0.05cm);
 %\draw[thick,dotted](4,2)--++(4,0);\draw[fill=black] (6,0) circle (0.05cm);
%\draw[decoration={zigzag},decorate](-2,4)--++(4,0); 
%\draw[blue,decoration={coil,segment length=0.5mm,amplitude=0.2mm},decorate] (4,2) arc (180:90:2);
%\draw[red,decoration={coil,segment length=0.5mm,amplitude=0.2mm},decorate] (8,2) arc (0:90:2); 
%\draw[fill=black] (6,4) circle (0.05cm);
%\tikz[every to/.style={bend right}]
% \draw[every to/.style={bend right}] (-2,0) to (-2,4);
\draw[thick] (-2,0) .. controls (-1.75,2) .. (-2,4);
% \draw[every to/.style={bend left}] (2,0) to (2,4);
\draw[thick] (2.15,0) .. controls (1.9,2) .. (2.15,4);
\end{tikzpicture}
\end{center}
\vskip-0.3cm 
\caption{\small The left and the right cutoff trajectories are illustrated 
as curves near the boundaries in the Penrose diagram of a typical 
two-sided black hole spacetime. 
The red and blue lines denote the original $\mu=\pm \pi/2$ boundaries of AdS$_2$. }\label{fig1}
\end{figure} 

}
\begin{figure}[htbp]  
\vskip0.3cm 
\begin{center}
\begin{tikzpicture}[scale=1.1]
%\draw[thick,blue %,-<
%](-2,0)--++(0,2);
%\draw[thick,blue] (-2,2)%node[left]{\large$t_l$}
%--++(0,2);
\draw%[thick]
(-3,0)--++(0,3);
\draw%[thick]
(-1.3,0)--++(0,3);
\draw%[thick]
(1.7,0)--++(0,3);
\draw%[thick]
(3.4,0)--++(0,3);
%\draw[blue,decoration={coil,segment length=0.5mm,amplitude=0.15mm},decorate] (-2,2) arc (180:270:2);\draw[red,decoration={coil,segment length=0.5mm,amplitude=0.15mm},decorate] (2,2) arc (0:-90:2); \draw[fill=black] (0,0) circle (0.05cm);
%\draw[decoration={zigzag},decorate](-2,0)--(2,0); 
% \draw[thick,dotted](-2,2)--++(4,0);
%\draw[decoration={zigzag},decorate](-2,4)--++(4,0); 
%\draw[blue,decoration={coil,segment length=0.5mm,amplitude=0.15mm},decorate] (6,4) arc (90:270:2);%\draw[red,decoration={coil,segment length=0.5mm,amplitude=0.15mm},decorate] (6,4) arc (90:-90:2); %\draw[fill=black] (6,0) circle (0.05cm);
 %\draw[thick,dotted](4,2)--++(4,0);\draw[fill=black] (6,0) circle (0.05cm);
%\draw[decoration={zigzag},decorate](-2,4)--++(4,0); 
%\draw[blue,decoration={coil,segment length=0.5mm,amplitude=0.2mm},decorate] (4,2) arc (180:90:2);
%\draw[red,decoration={coil,segment length=0.5mm,amplitude=0.2mm},decorate] (8,2) arc (0:90:2); 
%\draw[fill=black] (6,4) circle (0.05cm);
%\tikz[every to/.style={bend right}]
% \draw[every to/.style={bend right}] (-2,0) to (-2,4);
\draw[thick] 
(1.71,0.7) .. controls (1.8,1.5) .. (1.71,2.3);
\draw[thick] 
(3.39,0.7) .. controls (3.3,1.5) .. (3.39,2.3);
\end{tikzpicture}
\end{center}
\vskip-0.3cm 
\caption{\small On the left we draw the global AdS$_2$ as a strip where the left and right lines
denote the $\mu=- \pi/2, \pi/2$ boundaries of AdS$_2$, respectively.  On the right we illustrate the left and right cutoff trajectories 
as curves near the AdS$_2$ boundaries. }\label{fig1}
\end{figure}
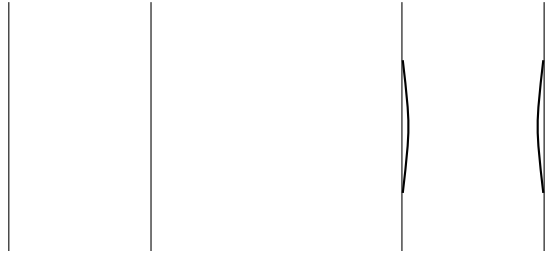

%In the context of nearly AdS$_{2}$ holography, we introduce , respectively. And,  the prescriptions for  metric and dilaton to get the cutoff boundaries are
As is well known, in this 2d gravity theory, there are no local dynamical gravity degrees of freedom in the bulk and all the pure gravity  dynamics are fully reflected in the boundary fluctuations of cutoff trajectories of AdS$_2$.  For this, one introduces the cutoff trajectories $(\tau_{r/l}(u),  \mu_{r/l}(u) )$  parametrized by  the  boundary time $u$  for the right and left cutoff boundaries. See the right diagram of Figure \ref{fig1}. The prescription for  metric and dilaton to get the cutoff boundary becomes
\begin{equation} \label{dscutoff}
ds^2|_{\text{cutoff}} = -\frac1{\epsilon^2} du^2\,, \qquad
\phi|_{\text{cutoff}} =  \frac{\bar\phi}{\epsilon}\,,
\end{equation}
%
%We will adopt the convention that the right boundary time $t_r$ runs upwards
%whereas the left boundary time $t_l$ runs downward. In other words, we identify
%\begin{equation} \label{bdryu}
%u=t_r = -t_l.
%\end{equation}
%See Section \ref{sec4} for the details.
and %Without bulk matters, 
the corresponding boundary dynamics  may  be identified as a combination of Schwarzian 
theories~\cite{Maldacena:2016upp,Jensen:2016pah, Engelsoy:2016xyb},
\begin{equation}  \label{schwarzian}
S = \int du\,  L_{r} +  \int du\, L_{l}\,,   \quad \ \ 
L_{r/l} =    \frac{\C}{2}\bigg[ \Big(\frac{\tau''_{r/l}}{\tau'_{r/l}}\Big)^{2} -\tau'^{2}_{r/l} \bigg]\,,   
\end{equation}
where  the total derivative terms are dropped  and the coupling $\C$ may be identified with  $\bar{\phi}$ that appears in the vacuum solution \eqref{dilatonvac}. 
% of the dilaton . 

For each Schwarzian Lagrangian, 
one may follow a standard procedure in higher derivative theory by
adding  Lagrange multiplier terms $ p_{\tau_{r/l}} ( \tau'_{r/l} - e^{\chi_{r/l}}/\C )$ to the Lagrangian where $p_{\tau_{r/l}}$ work as  Lagrange multipliers at this stage. Using the multiplier equations of motion, the above may be rewritten as
\begin{equation} \label{}
L_{r/l} = \frac{\C}{2}\chi'^2_{r/l} - \frac{1}{2\C} e^{2\chi_{r/l}}+ p_{\tau_{r/l}} \left( \tau'_{r/l} - \frac{1}{\C}e^{\chi_{r/l}} \right)\,.
\end{equation}
%%\cite{Jafferis:2019wkd}
By a further Legendre transform with  canonical momenta  $p_{\chi_{r/l}}$ %canonically
 conjugated to $\chi_{r/l}$, one  finds  %relevant canonical momenta can be introduced as 
\begin{align}    \label{}
L_{r/l} =  p_{\tau_{r/l}}  \tau'_{r/l} +p_{\chi_{r/l}}  \chi'_{r/l} -H_{r/l} \,,   
\end{align}
with the following left and right Hamiltonians \cite{Harlow:2018tqv,Jafferis:2019wkd,Bagrets:2016cdf}
\begin{equation} \label{bdhamiltonian}
 H_{r/l} = \frac{1}{2\C}  \Big[ p_{\chi_{r/l}}^{2} + 2 p_{\tau_{r/l}}\, e^{\chi_{r/l}}+  e^{2\chi_{r/l}}  \Big]\,.
\end{equation}
%
%\begin{align}    \label{}
%p_{\tau} &=  - \Big( e^{\chi} -C^{2} e^{-\chi} \chi'' \Big)\,,    \nonumber \\
%p_{\chi} &=  C \chi' \,. \nonumber 
%\end{align}
%
%Then, Hamiltonian becomes\footnote{The linear dependence of $H$ in $p_{\tau}$ tells us that $H$ is not bounded from below, which may be thought as an indication of instability of the system. This is,  of course, the usual aspect of higher derivative theory. However, in our two-sided case there would be a gauge symmetry which ensures the total Hamiltonian becomes positive on physical Hilbert space~\cite{Jafferis:2019wkd,Penington:2023dql}.}
%
%\begin{equation} \label{}
%H = \frac{1}{2C}  \Big[ p_{\chi}^{2} +  e^{2\chi} + 2 p_{\tau}\, e^{\chi} \Big]\,.
%\end{equation}
%
The linear dependence of $H_{r/l}$ in $p_{\tau_{r/l}}$ tells us that $H_{r/l}$  are  not bounded from below, which may be viewed as an indication of instability of the system. This is,  of course, the well-known aspect of higher derivative theory. However, in the present %two-sided 
case, there would be a gauge symmetry described in detail  below, which ensures the total Hamiltonian becomes positive on physical Hilbert space~\cite{Jafferis:2019wkd,Penington:2023dql}.

As reviewed in \cite{Lin:2019qwu},  the AdS$_2$ space has an  SL$(2,{\bf R})$ symmetry
under the isometric coordinate transformations that are generated by Killing vectors
\begin{align}    \label{}
\xi_{1} &= -\partial_\tau\,,   \nonumber \\
\xi_{2}&= - \cos\tau \sin\mu\, \partial_\tau - \sin\tau\cos\mu\, \partial_\mu \,,  \nonumber  \\
\xi_{3} &=   -\sin\tau \sin\mu\, \partial_\tau + \cos\tau\cos\mu\, \partial_\mu \,. %\nonumber 
\end{align}
Each of the above left-right boundary systems then possesses SL$(2,{\bf R})$ symmetry under the transformations that are induced by the bulk SL$(2,{\bf R})$ transformations along the left-right cutoff boundaries.
%Each of the above left right systems enjoys SL$(2,{\bf R})$ symmetry whose generators are given by
By the standard Noether procedure, the corresponding (quantum) SL$(2,{\bf R})$ generators may be 
constructed as \cite{Jafferis:2019wkd}
\begin{align}    \label{rlcharges}
J^{r/l}_{1} &= p_{\tau_{r/l}}\,,  \nonumber \\
J^{r/l}_{2} &=  \pm e^{\chi_{r/l}} \cos \tau_{r/l} \mp \sin \tau_{r/l}\, p_{\chi_{r/l}} \pm \cos \tau_{r/l} \, p_{\tau_{r/l}} \pm  \frac{i}{2}\sin\tau_{r/l}\,,  \nonumber  \\
J^{r/l}_{3} &=  \pm e^{\chi_{r/l}} \sin \tau_{r/l}  \pm \cos \tau_{r/l}\, p_{\chi_{r/l}} \pm \sin \tau_{r/l} \, p_{\tau_{r/l}} \mp \frac{i}{2}\cos\tau_{r/l}\,,  %\nonumber  
\end{align}
where the upper/lower signs are for the right/left quantities respectively.
These generators satisfy the SL$(2,{\bf R})$ algebra,
%
%\begin{equation} \label{}
$[J^{r/l}_{i},J^{r/l}_{j}]= i\epsilon_{ijk} \eta^{kl}J^{r/l}_{l}\ng,$
%\end{equation}
%
where $\epsilon_{ijk}$ is a totally antisymmetric symbol  with $\epsilon_{123} =1$
 and  $\eta^{ij}= \text{diag}(-1, 1,1)$. It is then straightforward to  check that 
\begin{equation} \label{}
 2\C H_{r/l} = \eta^{ij}J^{r/l}_{i}J^{r/l}_{j} -\frac{1}{4}
\,,
\end{equation}
which corresponds to the quadratic Casimir of SL$(2,{\bf R})$ and so ensures the SL$(2,{\bf R})$ invariance of the Hamiltonians.
%under the  % $J^{r/l}_{i}$  transformations.  
%The simultaneous eigenfunction of  $2CH$ and $J_{1}$  with eigenvalues of $\frac{s^{2}}{2}$ and $m$, respectively, is given by Whittaker W-function as 
%
%\begin{equation} \label{}
%\Psi_{m,s} (\tau, \chi)=  \frac{e^{im\tau}}{\sqrt{2\pi}} \frac{\sqrt{s\sinh 2\pi s} }{\sqrt{2\pi}} \Big|\Gamma\Big( m+\frac{1}{2} + is \Big)  \Big|e^{-\frac{\chi}{2}}W_{-m, is} (2e^{\chi})\,.
%\end{equation}
%

Now, by turning on the bulk matter field, the corresponding boundary flux along the cutoff boundaries may be in general nonvanishing and the equations of motion along the boundaries are modified as \cite{Maldacena:2016upp}
\begin{equation} \label{}
\C \left\{  \tan \frac{\tau_{r/l}(u)}{2} , u  \right\}' = \tau'^{2}_{r/l} ~  T_{\tau \mu}\big|_{r/l}\,,
\end{equation}
where   the Schwarzian derivative is defined by $\{ f(u),u\}\equiv  -1/2 (f''/f')^2+(f''/f')'$ and  the last $r/l$ denote the evaluation of the stress tensor at the right/left cutoff boundary, respectively. Thus, with this nonvanishing boundary flux, the boundary Lagrangians have to be modified accordingly through explicit coupling to the bulk matter field along the cutoff trajectories. In the context of the AdS/CFT correspondence, however, one  imposes the vanishing boundary condition
\begin{equation} \label{vbc}
\varphi \big|_{r/l} ={\cal O}(\cos^\Delta \ng\mu_{r/l})={\cal O}(\epsilon^\Delta)
\end{equation}
%
%which corresponds to the so-called vev deformation.
where $\Delta$ denotes the dimension of the operator dual to the bulk matter field. 
%We shall refine this boundary condition later on and  
In this paper, we shall limit our consideration to  the matter field with the vanishing boundary condition (\ref{vbc}) whose details will be further provided below. One is  led to the vanishing boundary flux along the cutoff trajectories in the $\eps\rightarrow 0$ limit. Then the forms of the boundary Hamiltonians 
in (\ref{bdhamiltonian}) remain intact while the effect of the bulk matter on the boundary 
systems are implicit through the constraints of the total conserved charges. With %the boundary condition 
(\ref{vbc}), the corresponding  bulk matter charges may be evaluated as
\be \label{mcharge}
J^m_i =-\int^{\pi/2}_{-\pi/2} d\mu~\sqrt{-g}\,  T^{\tau}_{~a}\, \xi^a_i = \int^{\pi/2}_{-\pi/2} d\mu~  T_{\tau a}\, \xi^a_i \,,
\ee
which are conserved and satisfy the SL$(2,{\bf R})$ algebra $[J^{m}_{i},J^{m}_{j}]= i\epsilon_{ijk} \eta^{kl}J^{m}_{l}$. In fact, there is an $\widetilde{\mbox{SL}}$$(2,{\bf R})%_{\mbox{\tiny G}}
$ gauge symmetry  generated by %transformation 
\be \label{gauges}
\tilde{J}_i %^{\mbox{\tiny G}}_i 
= J^r_i + J^l_i + J^m_i \,,
\ee
which leaves the full geometry, including
the cutoff boundaries, invariant. %This gauge symmetry is generated by 
Classically, one has the corresponding constraints, $\tilde{J}_i =0$, and imposing these %constraints 
leads to consistent solutions of the left-right boundary dynamics
once the bulk matter charges $J^m_i$ are specified appropriately. These boundary descriptions 
agree with those of the bulk gravity description, as was explicitly verified in \cite{Bak:2021qbo} for the case of $m^2=0$.
Quantum mechanically, %On the other hand, 
 imposing the constraints  properly on the wave function $\Psi$ 
\begin{equation} \label{}
\tilde{J}_i \, \Psi = 0
\end{equation}
will be the main part of our quantization of the system, whose details  will be discussed in the next section. 

For the resulting physical Hilbert space, we shall further impose 
$|\tau_{r}(u_1) - \tau_{l}(u_2) |  < \pi$ at any real $u_{1}, u_2$ for any nonvanishing $\Psi$.
%. % as our definition of the gravity theory. 
This condition basically ensures the causality constraint as the left-right boundary systems are causally disconnected from each other through the bulk\footnote{One can explicitly verify that this causality constraint is indeed respected  for full general bulk solutions with $m^2=0$ \cite{Bak:2021qbo}. }. In fact one may show that the above condition automatically follows from the condition  $|\tau_{r}(u_1) - \tau_{l}(u_2) | < \pi$ at some 
$u_1$ and $u_2$, say $u_1=u_2=0$ \cite{Penington:2023dql}.

\section{Canonical quantization}\label{sec3}
In this section, we consider the canonical quantization of JT theory without matter or with matter. The
presentation in this section is mostly a review of the construction given in  \cite{Penington:2023dql}.
Our starting point is the unconstrained Hilbert space ${\cal H}_0= L^2\otimes {\cal H}_m$ where
an $L^2$ function is further specified as a complex function of the variables $\tau_r, \tau_l, \chi_r, \chi_l$ that has a support only when
$|\tau_r -\tau_l| <\pi$. Here in the presence of matter, the function is dependent upon the matter 
part but we shall not spell out  this matter dependence explicitly in this section. 

Let us  impose the gauge constraint at quantum level. Since the $\widetilde{\mbox{SL}}$$ (2,{\bf R})$
%_{\mbox{\tiny G}}$
is noncompact, we may use a quantization scheme based on the equivalent classes defined by~\cite{Marolf:2008hg}
\be \label{}
\Psi \cong g \Psi
\ee
where $g \in \widetilde{\mbox{SL}}$$ (2,{\bf R})$. These equivalent classes are called the coinvariant 
classes of the group $\widetilde{\mbox{SL}}$$ (2,{\bf R})$. We then introduce inner product by the integral
\be \label{}
\langle \tilde{\Psi}| \Psi \rangle =\int dg \, (\tilde{\Psi}, \, g\Psi)\,,
\ee
where $dg$ is the left and right invariant measure of the group $\widetilde{\mbox{SL}}$$ (2,{\bf R})$.
It is clear that this inner product depends only on the equivalent classes of $\Psi$ and $\tilde{\Psi}$, so the formula 
defines the Hilbert space of coinvariants.  It also ensures the constraints
\be \label{constraints}
\tilde{J}_i\ng \int\ng dg \, g \Psi \cong 0
\ee  
at the quantum level, or equivalently $\langle \tilde{\Psi}|\, \tilde{J}_i\,| \Psi \rangle=0$ for any
choice of $| \Psi \rangle$ and % with a given 
$| \tilde\Psi \rangle$. 
Now we note that, for any $(\tau_r, \tau_l, \chi_r, \chi_l)$ with $|\tau_r -\tau_l| <\pi$, one may set $\tau_l = \tau_r=0$ and $\chi_r=\chi_l$ by an appropriate gauge transformation. This implies that the physical Hilbert space of coinvariants is generated by a gauge-fixed wavefunction of the form  \cite{Penington:2023dql}
 \be \label{}
\Psi = \delta(\tau_r)\delta(\tau_l)\delta(\chi_{rel}) \psi(\chi)
\ee
where $\chi \equiv \frac{1}{2} (\chi_r+\chi_l)$ and $\chi_{rel} \equiv \chi_r-\chi_l$. With this gauge-fixing condition, the constraints (\ref{constraints}) near $g=1$ are realized as
\begin{align}    \label{}
\tilde{J}_{1} &\cong  p_{\tau_r}+ p_{\tau_l}+J^m_1\cong 0\,,   \nonumber \\
\tilde{J}_{2} &\cong  p_{\tau_r}- p_{\tau_l}+J^m_2\cong 0\,,  \nonumber  \\
\tilde{J}_{3} &\cong  p_{\chi_r}- p_{\chi_l}+J^m_3\cong 0\,, % \nonumber 
\end{align}
and the inner product is reduced to
\be \label{}
\langle \tilde{\Psi}| \Psi \rangle =\int d\chi \, \tilde\psi^* (\chi) \psi(\chi)\,.
\ee
With help of the above relations, one may replace $ p_{\tau_{r/l}}$ and $ p_{\chi_{r/l}}$ %and $\chi_{r/l}$ 
by
\begin{align}    \label{}
p_{\tau_{r/l}} \cong  -\frac{1}{2}\left(J_1^m \pm J_2^m\right)\,,  % \nonumber \\
\ \ \ \  p_{\chi_{r/l}} =  \frac{1}{2}p_{\chi}\pm p_{\chi_{rel}} \cong \frac{1}{2}\left(p_\chi \mp J_3^m\right)\,,  %\nonumber 
%\chi_{r/l} &\cong  \chi\,,  \nonumber 
\end{align}
where the replacements are acting on the physical Hilbert space ${\cal H}=L^2(\chi)\otimes {\cal H}_m$.
Then $H_{r/l}$ %the reduced (physical) Hilbert space $L^2(\chi)\otimes {\cal H}_m$ 
become\cite{Penington:2023dql}
\be 
{2\C} H_{r/l} =  \frac{1}{4}\left(p_\chi \mp J_3^m\right)^2 - \left(J_1^m \pm J_2^m\right)\, e^{\chi}+  e^{2\chi} \,,
\ee
which are  acting on the physical Hilbert space ${\cal H}$ with $p_\chi =- i \partial_\chi$ satisfying  $[\chi, \, p_\chi ]=i$.
It is straightforward to show that $H_r$ commutes with $H_l$. %$[H_r, H_l]=0$. % in general. 
 In case of  pure JT gravity, one finds that $H_r$ is identical to
$H_l$ with the well known expression 
${2\C} H^0_{r/l}=\frac{1}{4}p_\chi^2  + e^{2\chi}$ \cite{Harlow:2018tqv}. In this case, 
the boost generator $H_r-H_l$ becomes  zero and merely induces  
 a  pure gauge transformation. % of the system.

\section{Explicit quantization % JT Gravity %Explicit construction 
with $m^2=0$}\label{sec4}

From now on, we shall consider  JT gravity with a massless field to be specific. For this %massless 
case, the bulk scalar field is dual to a dimension one ($\Delta=1$) 
operator in the boundary side. The corresponding bulk field may be solved by\footnote{The most general solutions for arbitrary $m$ are presented in \cite{Bak:2021qbo,Bak:2022cnz}. See also \cite{Spradlin:1999bn,Bak:2018txn}.} 
\be \label{scalarf}
\varphi=\sum^\infty_{n=1} \frac{1}{\sqrt{n \pi }}\sin n \big(\mu+\mbox{$\frac{\pi}{2}$}\big)\left(
a_n e^{-in \tau} +a^\dagger_n e^{-in \tau}\right)\,,
\ee
with  the boundary condition (\ref{vbc}).
%which is subject to the boundary condition (\ref{vbc}). 
Upon quantization, the creation and annihilation operators satisfy 
\be 
[a_m, a^\dagger_n]=\delta_{mn}\,,
\ee
while all the remaining commutators among them vanish identically. 
%From the definition of 
Starting with (\ref{scalarf}),
 the matter
 SL$(2,{\bf R})$ charges can be identified as
\begin{align}    \label{chragenn}
{J}^m_{1} &= - \sum^\infty_{n=1} n \,a^\dagger_n a_n \,,   \nonumber \\
{J}^m_{2} &=\frac{1}{2}\sum^\infty_{n=1}\mbox{\small$\sqrt{n(n+1)}$} \left(a^\dagger_n a_{n+1}+ a_n a^\dagger_{n+1} \right)\,,  \nonumber  \\
{J}^m_{3} &=\ng\frac{1}{2i}\sum^\infty_{n=1}\mbox{\small$\sqrt{n(n+1)}$} \left(a^\dagger_n a_{n+1}- a_n a^\dagger_{n+1} \right)\,, % \nonumber 
\end{align}
which follows from the definition of %SL$(2,{\bf R})$ 
charges in (\ref{mcharge}). Here, $J_1^m$ may 
involve an extra constant term but we fix this contribution to zero as it is required in order to allow a trivial representation on the vacuum sector of the bulk matter field. Of course, this vacuum sector should correspond to the pure JT theory. %The Casimir operator reads $C_m={J}^m_{i}{J}^m_{j}\eta^{ij}$ and 
Let us
introduce a %n oscillator 
 number operator defined by
%\be \label{}
$N_m=\sum^\infty_{n=1} %\mbox{$\sum^\infty_{n=1}$}
a^\dagger_n a_n$, %\,,
%\ee
which may be shown to be commuting with $J^m_i$. Then one has $[N_m, C_m]=0$, where $C_m$ denotes the Casimir operator given by
$\eta^{ij}{J}^m_{i}{J}^m_{j}$. It then follows that\footnote{Since the number operator and the Casimir do not commute with the boundary matter operator $\hat{\varphi}_{l/r}$, they are not a center of the left or right algebra ${\cal A}_{l/r}$.}
\be
[N_m, H_{r/l}]=[C_m, H_{r/l}]=0\,.
\ee
Hence, the time evolution of the system occurs within a sector with a given total oscillator number. The 
matter Hilbert space ${\cal H}_m$  has a basis
\be 
|\vec{k}\rangle =|k_1 k_2 k_3 \cdots \,\rangle \,, %\ \ \ a^\dagger_n a_n|\vec{k}\rangle= k_n |\vec{k}\rangle
\ee
with $a^\dagger_n a_n|\vec{k}\rangle= k_n |\vec{k}\rangle$ where $k_n$ is a nonnegative integer.
With this basis, one has
\be \label{nj1}
N_m|\vec{k}\rangle = \sum^\infty_{n=1} k_n |\vec{k}\rangle \,,  \ \ \ 
J_1^m|\vec{k}\rangle = -\sum^\infty_{n=1}n k_n |\vec{k}\rangle \,,%\ \ \ a^\dagger_n a_n|\vec{k}\rangle= k_n |\vec{k}\rangle
\ee
which in particular shows that the generator $J_1^m$ is nonpositive definite. The SL(2,{\,\bf R})-invariant matter vacuum state belongs to the trivial representation of SL(2,{\,\bf R}).
It is also  clear that the matter part of Hilbert space above the vacuum is given by a direct sum of the discrete series representation $D^-_{j=q} $ of SL(2,{\,\bf R}) specified with $C_m=q(1-q)$ and a given $N_m$.
See appendix \ref{AppA} for some details of representations of matter charges. 

Let us introduce the dual boundary operators which may be inserted along the left and right 
cutoff trajectories. In the present case of $\Delta=1$, the boundary operators may be identified
as
\be 
\hat{\varphi}_{r/l}=\lim_{\eps\rightarrow 0} \left(\frac{\bar\phi}{\epsilon}\right)^\Delta
\varphi\,\Big|_{r/l}\,.
\ee
In fact, this is a slight generalization of the standard AdS/CFT dictionaries reviewed in \cite{Harlow:2014yka}. In two dimensions, the cutoff trajectories become dynamical and one needs to take into account of these dynamical fluctuations of boundary geometries.   
   Using the relation %prescription (\ref{dscutoff}) 
%\be 
$\cos \mu \big|_{r/l} =\epsilon \tau'_{r/l} =\frac{\epsilon}{\C} e^{\chi_{r/l}}$,
%\ee
one finds
\be
\hat{\varphi}_{r/l}=\frac{e^{\chi_{r/l}}}{\sqrt{\pi}}\sum^\infty_{n=1}(\mp)^{n+1}\sqrt{n}
\left(
a_n e^{-in \tau_{r/l}} +a^\dagger_n e^{in \tau_{r/l}}\right)\,.
\ee
We note that
\be
[\hat{\varphi}_{r},\hat{\varphi}_{l}]=\frac{2i}{\pi}e^{\chi_r+\chi_l}\sum^\infty_{n=1}(-)^{n+1}n \sin(\tau_r\ng-\ng\tau_l)= 2ie^{\chi_r+\chi_l}\left(
\delta'(\tau_r\ng-\ng\tau_l\ng+\ng\pi)+\delta'(\tau_r\ng-\ng\tau_l\ng-\ng\pi)
\right)
\ee
where the second equality is defined over the interval $\tau_r-\tau_l \in [-\pi,\pi]$. Since 
$|\tau_r-\tau_l|<\pi$, one has $[\hat{\varphi}_{r},\hat{\varphi}_{l}]=0$. It is also straightforward to check that $[\tilde{J}_i ,\hat{\varphi}_{r/l}]=0$, so the left and right boundary operators are gauge invariant. Expressed in the physical Hilbert space variables after the gauge-fixing, these boundary operators become
\be
\hat{\varphi}_{r/l}=\frac{e^{\chi}}{\sqrt{\pi}}\sum^\infty_{n=1}(\mp)^{n+1}\sqrt{n}
\left(
a_n +a^\dagger_n \right)\,.
\ee
Finally, one may also check that $[H_{r/l} ,\hat{\varphi}_{l/r}]=0$. Any left-side operators
that are constructed out of $H_l$ and $\hat{\varphi}_l$ are commuting 
%with operators are commuting 
with the right-side operators that are generated by combinations of $H_r$ and $\hat{\varphi}_r$. 

There are two types of time evolutions in our theory. One is our system time defined for
the action (\ref{schwarzian}), where the time parameter $u$ evolves the left and right system equally %at the same time 
with the Hamiltonian
\be 
H_{tot}= H_r+H_l =\frac{1}{\C}\Big(\,\mbox{$\frac{1}{4}$}\left(p^2_\chi + {J_3^{\ng m}}^2\right) - J_1^m \, e^{\chi}+  e^{2\chi}\,\Big)\,.
\ee
In this note, we use this time evolution primarily.  One may introduce
a time-evolved operator by
\be
\hat{\varphi}_{r/l}(u) =e^{i u H_{tot}} \hat{\varphi}_{r/l} \, e^{-i u H_{tot}}
\ee
Alternatively, one may add an independent evolution by $H_{rel}=\frac{1}{2}(H_r-H_l)$ with a boost evolution parameter $u_{rel}$,  Unlike the case of pure JT, this boost operator becomes physical. Then one may evolve the left and right operators separately by
\be 
\hat{\varphi}_{r/l}(u_{r/l}) =e^{i u_{r/l} H_{r/l} }\hat{\varphi}_{r/l} \, e^{-i u_{r/l} H_{r/l}}\,.
\ee 
Since the left operators are fully commuting with the right operators, the two definitions agree
with each other. Thus, the latter evolution works equally well and is equivalent to the former in our case.
In either ways, %in 
the time-ordered $n (=n_r+n_l)$-point transition function %defined by
\be
G({u^r_1, \cdots, u^r_{n_r}; u^l_1, \cdots, u^l_{n_l}})=\langle \widetilde{\Psi}| {\cal T}_r\ng \prod^{n_r}_{k=1} \hat{\varphi}_{r}(u^r_{k}) \,{\cal T}_l\ng\prod^{n_l}_{k'=1} \hat{\varphi}_{l}(u^l_{k'})|\Psi\rangle\,,
%\rangle \ \ \ (k_{r/l}=1,2, \cdots, n_{r/l})
\ee
%is 
may be defined irrespective of
%independent of 
the orderings between the left and the right boundary operators. 
%does not matter at all. 
This reflects 
the fact that the left and  right cutoff boundaries are causally disconnected with each other. Note also that
$[N_m, \hat{\varphi}_{r/l}]\neq 0$ and $[C_m, \hat{\varphi}_{r/l}]\neq 0$. Hence with insertion of operators, the total number of oscillators and the value for the matter Casimir are not preserved in general. Of course, without insertion of extra operators, these two are preserved under the left and right Hamiltonian evolutions. 

\section{Comparison with bulk solutions}\label{sec5}
% Bulk classical description and Gauge-fixing.

% Bulk classical description and Gauge-fixing.
In this section, we consider  the gauge constraints and the gauge-fixing of the Schwarzian theories in the classical limit. %, which are derived from   bulk solutions.  
%and compare them with those from  bulk solutions. 
First, one may recall the equations of motion given by the Hamiltonians~\eqref{bdhamiltonian} 
\begin{equation} \label{SchEOM}
p'_{\tau_{r/l}} =0\,,   \qquad %\C 
p_{\chi_{r/l}} = \C \chi'_{r/l}\,, \qquad %\C ( 
e^{2\chi_{r/l}} + p_{\tau_{r/l}} e^{\chi_{r/l}} = -\C p'_{\chi_{r/l}}\,,   \qquad 
 e^{\chi_{r/l}}  =\C \tau'_{r/l}\,, 
\end{equation}
which retain their forms even in the presence of the  matter. % in our setup. 

The solutions to these equations are given by
\begin{align}    \label{}
\frac{1}{\C} p_{\tau_{r/l}} &= \pm\,  C_{r/l}\,,   \\
\frac{1}{\C}  e^{\chi_{r/l}} &=  A_{r/l} \cos \tau_{r/l} + B_{r/l} \sin \tau_{r/l} \mp \,  C_{r/l} \,,  \label{echisol} \\
\frac{1}{\C}  p_{\chi_{r/l}} & = -A_{r/l} \sin \tau_{r/l} + B_{r/l} \cos \tau_{r/l} \,,
\end{align}
where $A_{r/l}, B_{r/l}$ and $C_{r/l}$ are  integration constants. The last equation in~\eqref{SchEOM},  $e^{\chi_{r/l}}  = \C \tau'_{r/l}$, together with the  solution~\eqref{echisol} leads to  boundary cutoff trajectories~\cite{Bak:2021qbo}  parametrized as
\begin{equation} \label{cutofftra}
\tanh \mbox{$\frac{1}{2}$}L_{r/l}(u-u_{0}^{r/l}) = \textstyle{\sqrt{\frac{1+q_{r/l} }{1-q_{r/l} }} }~ \tan \frac{1}{2}\Big(\tau_{r/l}(u) -\tau^{r/l}_{B}\Big) \,,\end{equation}
where $u^{r/l}_{0}$ are  another integration constants and\footnote{In the pure JT gravity, $L_{r/l}$ reduces to the horizon radius $L$   since $A_{r}=A_{l}= L $ for the vacuum black hole solution in~\eqref{dilatonvac}. This will become clearer after we discuss the match between the bulk solution and the boundary Schwarzian solution provided below.} 
\begin{equation} \label{}
\tan \tau^{r/l}_{B} \equiv  \mbox{$\frac{B_{r/l}}{A_{r/l}}$}\,, \qquad   q_{r/l} \equiv \mbox{$\frac{\pm C_{r/l}}{\sqrt{A_{r/l}^{2}+B_{r/l}^{2}}}$}\,, \qquad L_{r/l} \equiv \mbox{$\sqrt{A_{r/l}^{2} + B_{r/l}^{2} -C_{r/l}^{2}}$}\,.
\end{equation}
%
%Note that  $J_{i}^{r/l}$ in~\eqref{} act on the canonical variables as the SL$(2,{\bf R})$ generators under the Poisson bracket $\{~,\}_{P.B}$. For instance,  
%%
%\begin{equation} \label{}
%\{ J_{1}^{r/l}, e^{\chi_{r/l} }\} = 0\,, \qquad 
%\{ J_{2}^{r/l}, e^{\chi_{r/l}} \} = e^{\chi_{r/l} }\sin \tau_{r/l}\,, \qquad \{ J_{3}^{r/l}, e^{\chi_{r/l} }\} = - e^{\chi_{r/l} } \cos \tau_{r/l}\,. 
%\end{equation}
%
For these solutions, the on-shell values of SL$(2,{\bf R})$ generators $J_{i}^{r/l}$ in~\eqref{rlcharges} are given by 
\begin{align}    \label{Schargesol}
J_{1}^{r/l} \big|_{sol} = \pm\, \C\, C_{r/l}\,, \qquad J_{2}^{r/l}\big|_{sol} = \pm\, \C\,A_{r/l}\,, \qquad J_{3}^{r/l}\big|_{sol} = \pm\, \C\, B_{r/l}\,,
\end{align}
and so the on-shell values of the Hamiltonians are given by 
\begin{equation} \label{}
H_{r/l} \big|_{sol} = \frac{\C}{2}\big(A_{r/l}^{2} +B_{r/l}^{2} -C_{r/l}^{2}\big)\,.
\end{equation}

As is well-known as the Darboux's theorem, the solution space of the equations of motion  is symplectomorphic to the phase space  in classical mechanics~\cite{Crnkovic:1986ex,Wald:1995yp}. In  this regard, the  eight constants $A_{r/l}, B_{r/l}, C_{r/l}$ and $u_{0}^{r/l}$ describing classical solutions correspond to the eight-dimensional (unconstrained) phase space described by four variables $(\tau_{r/l}, \chi_{r/l})$ and their canonical conjugate momenta $(p_{\tau_{r/l}}, p_{\chi_{r/l}})$. To obtain  physical phase space  with the causality constraint $|\tau_r -\tau_l| <\pi$, we need to take a symplectic quotient by the constraint group $\widetilde{\mbox{SL}}$$ (2,{\bf R})$.   This quotient or reduction of variables can be understood as fixing the integration constants.  The constants $u_{0}^{r/l}$ may be chosen,  which corresponds to a certain gauge choice,  such as 
\begin{equation} \label{}
\tanh \mbox{$\frac{1}{2}$} L_{r/l}\,u_{0}^{r/l} = \textstyle{\sqrt{\frac{1+q_{r/l} }{1-q_{r/l} }} }~ \tan \frac{1}{2}\tau^{r/l}_{B} \,,
\end{equation}
and then, $\tau_{r/l}(u=0) =0$.  One may  note that the gauge invariant combination of remaining constants $A_{r/l}, B_{r/l}$ and $C_{r/l}$ appears in the cut-off trajectory expression in~\eqref{cutofftra},  which is given by $L_{r/l}$ or equivalently the right/left energies $H_{r/l}|_{sol}$. So, the solution space is described by the  variables $L_{r/l}$ ({\it i.e.}  $H_{r/l}|_{sol}$).  In pure JT gravity, $H_{r} + H_{l}= 2H_{r} = 2H_{l} $ and its conjugate variable $\frac{1}{2}(\tau_{r}+\tau_{l})$ form two-dimensional phase space~\cite{Harlow:2018tqv,Penington:2023dql}, while in JT gravity with matter $H_{r}+ H_{l}$ and $H_{r}-H_{l}$ give us different time evolutions and energies.  Of course, this reduction can also be understood from the canonical variables. Concretely, by using $\tilde{J}_{1}$ and $\tilde{J}_{2}$, one can set $\tau_{r} = \tau_{l} =0$. Using the remaining constraint generator  $\tilde{J}_{3}$, one can set $e^{\chi_{r}} = e^{\chi_{l}}$. See appendix~\ref{AppB} for the details of the gauge-fixing.

Now, let us consider the bulk scalar solution and its on-shell matter $J_{\, i}^{m}$ charges to check the gauge constraints on the classical solutions.  First of all, one may note that
  the on-shell matter charges, computed by the bulk integral~\eqref{mcharge} for  the classical scalar field solution $\varphi$ given in~\eqref{scalarf},   takes the same form with %
\eqref{chragenn}.
%\eqref{mcharge}
 This can be rewritten as
\begin{equation} \label{mchargesol}
J_{i}^{m}\big|_{sol} = -(Q_{i}^{r} -Q_{i}^{l})\,,
\end{equation}
where $Q^{r/l}_{i}$ read
\begin{align}    \label{Qexp}
Q_{1}^{r/l} &= \pm %\frac{1 }{2}
\sum_{n=1}^{\infty}\mbox{$\frac{1 }{2}$} n a^\dagger_{n}a_{n}\,, \nonumber \\
Q_{2}^{r/l} &= \bar{\phi}L \mp \sum_{n=1}^{\infty}\mbox{$\frac{1 }{4}$} \sqrt{n(n+1)}\,(a^{\dagger}_{n}a_{n+1} + a_{n+1}a^{\dagger}_{n} )\,,  \nonumber \\
Q_{3}^{r/l} &= \mp \sum_{n=1}^{\infty}\mbox{$\frac{1 }{4i}$}\sqrt{n(n+1)} \,(a^{\dagger}_{n}a_{n+1} - a^{\dagger}_{n+1}a_{n} ) 
\,,
\end{align}
where $a^{\dagger}_{n}$ should be interpreted as the complex conjugate of $a_{n}$ in the classical solutions. 

We have judiciously  rewritten the on-shell matter charges $J^{m}_{i}|_{sol}$ in terms of $Q^{r/l}_{i}$'s, since those are related to the asymptotic form of the bulk dilaton field $\phi$.
For the bulk scalar field solution, $\varphi_{sol}$ under a vanishing boundary condition, the bulk dilaton solution can be obtained by solving~\eqref{phieq}. 
 From the explicit asymptotic expressions of the bulk dilaton solution, one can see that the dilaton solution  $\phi$ at the cutoff boundaries  takes the form of the vacuum solution as (see Appendix~\ref{AppC} for a summary of these solutions in~\cite{Bak:2021qbo})
\begin{equation} \label{phicutoff}
\phi \underset{\mu\rightarrow \mu^{r/l}_{c}} {\longrightarrow}\eta^{ij}Q^{r/l}_{i}Y_{j}\Big|_{\mu^{r/l}_{c}}\,, \qquad Y_{i} = \Big( \tan\mu,\frac{\cos\tau}{\cos\mu}, \frac{\sin\tau}{\cos \mu}\Big)\,.
\end{equation}
Equivalently, using the equations of motion %given  in~
\eqref{phieq}, one immediately sees that the relevant bulk integral 
%for the bulk solution 
reduces to a surface term, resulting in the above expression of $J^{m}_{i}|_{sol}$. For instance, $J^{m}_{\,1}|_{sol}$ can be computed as
\begin{equation} \label{}
J^{m}_{1}|_{sol} = - \int^{\mu^{r}_{c}}_{\mu^{l}_{c}} d\mu~ T_{\tau\tau}|_{sol} = \frac{1}{\cos\mu} \frac{\partial}{\partial \mu}  (\phi\, \cos\mu) \Big|^{\mu^{r}_{c}}_{\mu^{l}_{c}} = - (Q^{r}_{1}-Q^{l}_{1})\,,
\end{equation}
which shows why the matter charges $J^{m}_{i}|_{sol}$ are related to the asymptotic forms of the dilaton field $\phi$. Note  that the left and right constants $Q^{r/l}_{i}$ are not independent but related by
\begin{equation} \label{}
Q^{r}_{1} = -Q^{l}_{1}\,, \qquad \bar{\phi}L -Q^{r}_{2} =-( \bar{\phi}L- Q^{l}_{2})\,, \qquad Q^{r}_{3} = -Q^{l}_{3}\,.
\end{equation}

Now,  let us check the gauge constraints by relating the bulk solutions to the boundary solutions in Schwarzian variables $(e^{\chi_{r/l}}, \tau_{r/l})$ through the cutoff conditions given in~\eqref{dscutoff}. By using the relation of the dilaton at the cutoff trajectories with the Schwarzian variables, one  obtains
\begin{equation} \label{}
e^{\chi_{r/l}} = \C \tau'_{r/l}   =  \frac{\C}{\epsilon} \cos\mu^{r/l}_{c}= \phi~\cos\mu^{r/l}_{c} |_{\mu\rightarrow \mu_{c}^{r/l}} = \mp Q^{r/l}_{1} + Q^{r/l}_{2}\cos\tau + Q^{r/l}_{3}\sin\tau\,,
\end{equation}
where the metric cutoff condition~\eqref{dscutoff} is used in the second equality, the dilaton cutoff condition~\eqref{dscutoff} with $\C =\bar{\phi}$ is used in the third equality, and the asymptotic form of $\phi$ in~\eqref{phicutoff} is used in the last equality. By matching this expression to the boundary solution in~\eqref{echisol}, one can deduce
that it is consistent with the gauge constraint $\tau_{r}=\tau_{l}$ by taking $\tau=\tau_{r/l}$, and   that 
\begin{equation} \label{}
\C\, C_{r/l} = Q^{r/l}_{1}\,, \qquad \C\, A_{r/l} = Q^{r/l}_{2}\,, \qquad \C\, B_{r/l} = Q^{r/l}_{3}\,.
\end{equation}
Using this matching of constants, \eqref{Schargesol} and \eqref{mchargesol}, it is straightforward to check   
\begin{equation} \label{}
\tilde{J}_{i}\big|_{sol} =\big( J^{r}_{i} + J^{l}_{i} + J^{m}_{i}\big) |_{sol}  = 0\,,
\end{equation}
which tells us that the gauge constraints are automatically satisfied in classical solutions when the Schwarzian variables and the dilaton at the cutoff trajectories are related by the cutoff conditions~\eqref{dscutoff}. 

As a side remark, we would like to note that matter charges are related to the left or right coefficients $Q^{r/l}_{i}$  as
\begin{equation} \label{}
 Q_{1}^{r/l} = \mp\,   \frac{1}{2}  J^{m}_{1}\big|_{sol}   \,, \qquad  Q_{2}^{r/l}  = \bar{\phi}L \mp  \frac{1}{2} J_{2}^{m}\big|_{sol} \,, \qquad Q_{3}^{r/l}   = \mp\,  \frac{1}{2}J_{3}^{m}\big|_{sol}   \,.
\end{equation}

%%
%Here, the upper/lower signs should be taken for the right/left sides, respectively. 
%%
%\begin{equation} \label{}
%\qquad Z^{r/l}_{i} = (C_{r/l},A_{r/l},B_{r/l})\,, 
%\end{equation}
%%

%For pure JT gravity, using a $\mbox{SL}(2, {\bf R})$ rotation, one can set $J_{1}^{r} = J_{3}^{r}=0$ and $J_{2}^{r} >0$ for the orbits with $\eta^{ij}J_{i}^{r}J_{j}^{r} >0$~\cite{Penington:2023dql}.    Then, the constraint, $J_{1}^{r}+J_{1}^{l}=0$ sets $J_{1}^{r}=J_{1}^{l} =0$, which lead to  $C_{r} = C_{l} =0$,  while  the constraint, $J_{3}^{r}+J_{3}^{l}=0$ does $B_{r} = B_{l} =0$.
%one can take,  
%%
%\begin{equation} \label{}
%e^{\chi_{r}} = 
%\end{equation}
%

\section{Two-sided correlation functions}\label{sec6}
\begin{figure}[htbp]   
\begin{center}
\vskip0.1cm
\begin{tikzpicture}[scale=1]
%\draw[blue, thick,-<](-2,0)--++(0,2);\draw[blue,thick] (-2,2)node[left]{\large$t_l$}--++(0,2);
%\draw[thick,red, ->](2,0)--++(0,2);\draw[thick,red] (2,2)node[right]{\large$ t_r$}--++(0,2);
%\draw[blue,decoration={coil,segment length=0.5mm,amplitude=0.15mm},decorate] (-2,2) arc (180:270:2);
\draw %[red,decoration={coil,segment length=0.5mm,amplitude=0.15mm},decorate] 
(2,2) arc (0:360:2); \draw[%red,
decoration={coil,segment length=6mm,amplitude=0.3mm},decorate] (-0.2,3.83) arc (98:262:1.83);
\draw[%red,
decoration={zigzag,
segment length=6mm,amplitude=0.4mm},decorate] (0.1,3.83) arc (86:-86:1.83);
\draw[red] (-0.2,3.83)%node[below,xshift=0.1cm]{$\chi_c$} 
arc (180:360:0.15);
\draw[red] (-0.2,0.21) arc (180:-5:0.15);
%\draw%[thick] 
%(-0.6,0.29) .. controls (-0.0,0.58).. (0.6,0.26);
%\draw[fill=black] (0,0) circle (0.05cm);
%\draw[decoration={zigzag},decorate](-2,0)--(2,0); 
% \draw[thick,dotted](-2,2)--++(4,0);
%\draw[decoration={zigzag},decorate](-2,4)--++(4,0); 
%\draw[decoration={zigzag,segment length=1mm,amplitude=0.5mm},decorate](-2,4)--++(0.2,+0.35); 
%\draw[decoration={zigzag,segment length=1mm,amplitude=0.5mm},decorate](2,4)--++(-0.2,+0.35); 
%\draw[decoration={zigzag,segment length=1mm,amplitude=0.5mm},decorate](-2,0)--++(0.2,-0.35); 
%\draw[decoration={zigzag,segment length=1mm,amplitude=0.5mm},decorate](2,0)--++(-0.2,-0.35); 
%\draw[thick,brown,dotted] (-2,0)--++(4,4);
%\draw[thick,brown,dotted] (+2,0)--++(-4,4);
%\draw (-2,0)--++(4,4);
%\draw (+2,0)--++(-4,4);
\draw[fill,fill opacity=0.02]
%[fill=%fill=yellow!80!black,draw opacity=0.1]  
%[red,decoration={coil,segment length=0.5mm,amplitude=0.15mm},decorate] 
(8,2) arc (0:360:2); \draw[%red,
%fill,fill opacity=0.03,
decoration={coil,segment length=6mm,amplitude=0.3mm},decorate] (6.55,3.7) arc (74:286:1.80);
\draw[%red,
%fill,fill opacity=0.03,
decoration={zigzag,
segment length=5mm,amplitude=0.4mm},decorate] (6.85,3.58) arc (64:-65:1.75);
\draw[red%,fill=black,fill opacity=0.3
] (6.55,3.7)%node[below,xshift=0.1cm]{$\chi_c$} 
arc (160:338:0.16)%;
%\draw[red]
 (6.55,0.24) arc (200:45:0.17);
\draw[fill=black] (6.7,3.65) circle (0.07cm);
\draw[fill=black] (6.71,0.31) circle (0.07cm);
\draw%[red] 
(4.55,1.9) node%[right]
{\small $\beta_l$};\draw%[red] 
(7.55,1.9) node%[right]
{\small $\beta_r$};
\draw%[red] 
(-1.46,1.9) node%[right]
{\small $\beta_l$};\draw%[red] 
(1.51,1.9) node%[right]
{\small $\beta_r$};
%\draw[blue,decoration={coil,segment length=0.5mm,amplitude=0.15mm},decorate] (6,4) arc (90:270:2);\draw[red,decoration={coil,segment length=0.5mm,amplitude=0.15mm},decorate] (6,4) arc (90:-90:2); \draw[fill=black] (6,0) circle (0.05cm);
 %\draw[thick,dotted](4,2)--++(4,0);\draw[fill=black] (6,0) circle (0.05cm);
%\draw[decoration={zigzag},decorate](-2,4)--++(4,0); 
%\draw[blue,decoration={coil,segment length=0.5mm,amplitude=0.2mm},decorate] (4,2) arc (180:90:2);
%\draw[red,decoration={coil,segment length=0.5mm,amplitude=0.2mm},decorate] (8,2) arc (0:90:2); 
%\draw[fill=black] (6,4) circle (0.05cm);
\end{tikzpicture}
\vskip-0.1cm
\caption{\small On the left we draw the two-side evolution for the pure JT theory. On the right,
we depict the evolution for JT theory with matter. The big dots represent insertions of the boundary operators at the initial  and the final points of the evolution. The red curves in each diagram represent
the initial and final cutoff geodesics.
}
\label{Fig2}
\end{center}
\end{figure}
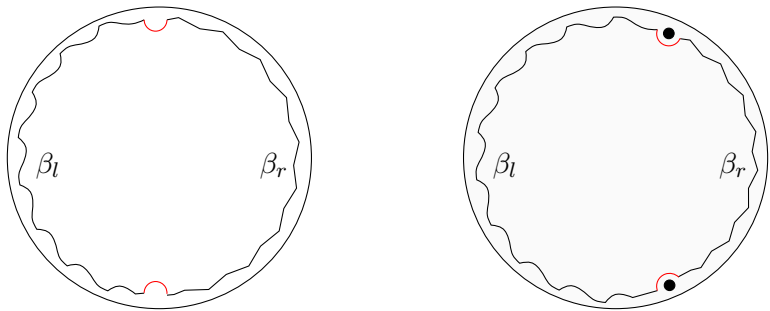

In this section, we shall consider the partition function and two-sided correlation functions of JT theory with  matter
from the viewpoint of two-sided picture.  Let us begin with the case of pure JT theory. 
Without matter contribution, the total Hamiltonian becomes
\be
H_{tot}= H_r+H_l = 2H_r =%\frac{1}{\C}
\big(\,\mbox{$\frac{1}{4}$}p^2_\chi +  e^{2\chi}\,\big)/\C\,.
\ee 
%It says that a particle is moving with a Liouville potential  
This is a Liouville quantum-mechanical system that involves an exponential potential.  
Note that  the renormalized geodesic length between two boundary points 
%specified by 
$\tau_l(u)$ and $\tau_r(u)$  is given by % is related to the variable $\chi$ by
\be 
\ell_{ren}\equiv \ell_{bare}-\ln 2\phi|_r -\ln 2\phi|_l =\ln \left(\ng
\frac{\cos^2 \frac{\tau_r-\tau_l}{2}}{\C^2\tau'_l \tau'_r} 
\ng\right)=-2\chi
\ee
where, for the first equality, (\ref{dscutoff}) is used and  the last equality follows from the gauge-fixing condition in the above. 
The corresponding eigenvalue problem, 
\be 
 H_{tot} \, \psi_s (\chi)= % s^2/\C 
\frac{s^2}{\C}  \,
\psi_s(\chi)\,,
\ee
can be solved by \cite{Harlow:2018tqv}
\begin{align}    \label{}
\psi_s (\chi)= N_s \, K_{2is} (2 e^\chi) \,,  % \nonumber \\
\ \ \ \ \  N_s =\frac{2}{\pi} (2s \sinh 2\pi s)^{\frac{1}{2}}\,, % \nonumber 
\end{align}
which satisfies the scattering normalization 
\be
\int^\infty_{-\infty} \ng d\chi\, \psi_s^*(\chi) \psi_{s'}(\chi) =\delta(s-s')\,.
\ee
%where $K_\nu (x)$ denotes the modified Bessel function of the second kind.
In the scattering regime of $\chi\rightarrow -\infty$, the wavefunction behaves as  %becomes
\be
\psi_s\rightarrow \frac{\phantom{a}\Gamma(-2is)}{\sqrt{\pi}\,|\Gamma(-2is)|}\,\big(\,e^{2is\chi}+R(s) e^{-2is\chi}\,\big)\,,
\ee
where the reflection amplitude may be identified as $R(s)=\frac{\Gamma(\,2is\,)}{\Gamma(\ng-\ng 2is)}$.  In the forbidden region of $\chi \rightarrow \infty$, on the other hand, the wavefunction decays doubly-exponentially as
\be
\psi_s(\chi)\rightarrow  N_s \,\sqrt{
\frac{{\pi}}{4e^{\chi}}}\, e^{-2 e^{\chi}}\,.
\ee
Now let us turn to the evaluation of the disk partition function in the two-sided picture. The relevant  density of states is basically one-sided quantity whereas our physical Hilbert space is inherently two-sided. In this respect, currently there is no well-defined procedure computing 
the disk partition function based on the two-sided description. Here we follow 
 the proposal in \cite{Lin:2022zxd} 
\be
Z(\beta) \propto %{\mbox{lim}}
\lim_{\chi^c \rightarrow \infty} \langle \chi^c |\, e^{-\frac{\beta}{2} H_{tot}} |\chi^c  \rangle\,,
\ee
which is based on the picture in the left side of Figure \ref{Fig2}. Since $H_l=H_r$ in the present case, %$\frac{1}{2}
$\beta H_{tot}/2$ may be replaced by $\beta_l H_l +\beta_r H_r$ with $\beta_l +\beta_r =\beta$.

% %

In this two-sided picture, one starts the evolution from an initial geodesic connecting two slightly separated boundary
points %around 
somewhere on the bottom side %boundary location 
as illustrated in the left side of Figure \ref{Fig2}. 
 Its renormalized length $\ell^c_{ren}\, (=-2 \chi^c)$ goes to  negative infinity as the above two points approach each other. Then we evolve through the bulk  leading to the final geodesic between two regularized points on the top side.
Basically the evolution is based on the propagator with an appropriate Boltzmann weight, which defines the path integral computation in the two-sided picture. 
With this prescribed regularization, one finds
\be
Z(\beta) \propto %{\mbox{lim}}
\lim_{\chi^c \rightarrow \infty} W(\chi^c) \int_{0}^\infty ds s \sinh 2\pi s\, e^{-\beta \frac{s^2}{2\C}}
\ee
where
\be 
W(\chi)=\frac{2\, e^{-4 e^{\chi}} }{\pi \, e^{\chi}}\,.
\ee
Note here that the factor $W(\chi^c)$ is independent of the variable $s$ and may be absorbed into 
the overall coefficient of the partition function or the constant part of the entropy $S_0$. Hence up 
to this overall coefficient, the disk partition function may be identified as
\be \label{JTpartition}
Z(\beta)= \int_{0}^\infty ds s \sinh 2\pi s\, 
e^{-\beta \frac{s^2}{2\C}}= \frac{\sqrt{2}\,\C^{\frac{3}{2}}}{\beta^{\frac{3}{2}}} e^{\frac{2\pi^2 \C}{\beta}}
\ee   
which agrees with the previous results based on the one-sided picture \cite{Stanford:2017thb,Cotler:2016fpe}  (See also~\cite{Mandal:2017thl,Mertens:2017mtv,Mertens:2018fds,Blommaert:2018oro} for related works).   The question yet remaining is how to specify
the initial (or final) state and an alternative based on the Hartle-Hawking state on a half disk
is given in \cite{Kolchmeyer:2023gwa}.

We now turn to the case of JT gravity with a  massless matter field. The two-sided function now depends 
on $\beta_r$ and $\beta_l$ since $H_r$ and $H_l$ differ from each other\footnote{$\beta_l$ and $\beta_r$ are simply  left and right Euclidean evolution parameters, which should not be confused  with the left and right temperatures. %of the two-sided system. 
%Also the $\beta$ parameter, in the presence of matter, should be interpreted as the total boundary length of the Euclidean disk.
}.  In the semiclassical regime,
the left and right black holes involving a nontrivial matter field indeed become different from each other as was constructed explicitly in \cite{Bak:2021qbo}. The corresponding Euclidean disk geometry becomes %inherently 
two-sided\footnote{%In three dimensions,  
The Euclidean geometry of the 3d Janus two-sided black hole was constructed in  \cite{Bak:2007qw},
%, which has a topology of a thermal disk $\times \, S^1$;
%The  
whose boundary of the thermal  disk part  
is  intrinsically two-sided involving different left and right Hamiltonians.
% that differ from each other.
} with insertion of operators  in the bottom and the top region.
% \cite{Bak:2022cnz}.
%In fact, the upper half of the disk geometry is fixed by the lower half of the disk geometry by the reflection positivity requirement. 
This insertion of matter state (as a linear combination of $|\vec{k}\rangle$)
%which 
induces a state at the initial curve as $|\Phi_{I}\rangle =\sum_{\vec{k}}\, %\psi_{{\vec{k}}} 
 \delta(\chi-\chi^c_{\vec{k}}) 
\, c_{{\vec{k}}}\, |\vec{k}\rangle $, by which   the bulk will be affected in general.  We also assume 
the final cutoff state $|\Phi_{F}\rangle=|\Phi_{I}\rangle$ for simplicity.
%based on the reflection positivity. (
Of course, this assumption can be relaxed and the definition may be generalized to the case where $|\Phi_{F}\rangle\neq|\Phi_{I}\rangle$.
With this preparation, we consider %our proposal is that
%\footnote{ Note that this is not a conventional partition function in a strict sense, but we shall call it simply partition function in this note.}
\be
Z_I(\beta_r, \beta_l) \propto %{\mbox{lim}}
\lim_{\chi^c_{\vec{k}} \rightarrow \infty} \langle \Phi_I |\, e^{-(\beta_r H_r + \beta_l H_l) }|\Phi_I  \rangle
\ee
%in our two-sided description.
where the way to send $\chi^c_{\vec{k}}$ to infinity will be specified further below. The corresponding two-sided evolution is depicted in Figure \ref{Fig2}. Again one begins with an initial geodesic curve
connecting two slightly separated points in the bottom region.  %While taking the limit, 
The precise locations of these two points 
may be adjusted by an infinitesimal amount depending  on 
each matter basis state %specified by 
$|\vec{k}\rangle$ (see below). From this, we evolve the two-sided  system with a Boltzmann weight 
$e^{-(\beta_r H_r + \beta_l H_l) }$  which ends  up with the final geodesic curve prescribed by the same way as the initial one. %In this evolution, one has 
The corresponding  left/right  evolution times are given by $\beta_l$/$\beta_r$ with the Hamiltonians $H_l$/$H_r$, respectively.  Due to the initial and final insertions of operators, the bulk state and the left  and right 
Hamiltonians are affected in general. In the semiclassical limit, this corresponds to  the so-called vev deformation whose details are studied in \cite{Bak:2022cnz}.   

To be specific, let us consider the case with $\beta_r=\beta_l=\frac{\beta}{2}$ and $|\Phi_{I}\rangle=\delta(\chi\ng-\ng\chi_{\vec{k}}^c)%|\chi^c_{\vec{k}}\rangle
\,|\vec{k}\rangle$. In this case, one has
$\beta_r H_r + \beta_l H_l  = \frac{\beta}{2}H_{tot}$. The relevant eigenvalue problem 
$ H_{tot}|\Phi\rangle =E |\Phi\rangle $ may be solved perturbatively by decomposing 
\be
H_{tot}=H_{(0)} +H_{(1)}
\ee
where
\be
H_{(0)}=\frac{1}{\C}\Big(\,\mbox{$\frac{1}{4}$} p^2_\chi - J_1^m \, e^{\chi}+  e^{2\chi}\,\Big)\,,
 \ \ \ \ H_{(1)}=\frac{1}{4\C} ({J^{ m}_3})^2\,.
\ee
We solve the zeroth-order eigenvalue problem $\C H_{(0)}|\Phi\rangle_{(0)}=s^2 
%\frac{s^2}{\C} 
|\Phi\rangle_{(0)}$ with a state of the form $|\Phi\rangle_{(0)}=\psi_{q,s}(\chi)\,|\vec{k}\rangle$. This leads to an eigenvalue equation
\be  \label{qeq}
 \Big(\,\mbox{$\frac{1}{4}$} p^2_\chi +q \, e^{\chi}+  e^{2\chi}\,\Big) \psi_{q,s}(\chi) =s^2 \psi_{q,s}(\chi)\,,
\ee
where $q=\sum^\infty_{n=1} n k_n \ge 0$. This $q$-dependent potential is everywhere nonnegative definite and 
becomes zero as $\chi\rightarrow -\infty$.
%The equation (\ref{qeq})  
This
problem is solved by
\begin{align}    \label{}
\psi_{q,s} = N_{q,s} \, Y_{q,s} (4 e^\chi) \,,   %\nonumber \\
 \ \ \ \ N_{q,s} =\frac{2}{\pi} (2s \sinh 2\pi s)^{\frac{1}{2}}\Big|
\frac{\Gamma(\frac{1}{2}+q+2si)}{\Gamma(\frac{1}{2}+2si)}
\Big| \,, % \nonumbe
\end{align}
with $Y_{q,s}(z)=
\sqrt{%\frac{\pi}{z}%
\pi/z
} 
%\sqrt{\frac{\pi}{z}}
\, W_{\ng- q,2si}(z)$ where $W_{\kappa,\mu}(z)$ is the 
Whittaker function satisfying
\be
\mbox{$\frac{d^2}{dz^2}$} W_{\kappa,\mu} +\big(\ng-\ng \mbox{$\frac{1}{4}$}+q z^{-1}+(\mbox{$\frac{1}{4}$}-\mu^2)z^{-2} \big)W_{\kappa,\mu}=0\,.
\ee
The wavefunction %again 
is again %satisfies
%the 
%scattering normalization 
scattering-normalized as
\be
\int^\infty_{-\infty} \ng d\chi\, \psi_{q,s}^*(\chi) \, \psi_{q,s'}(\chi) =\delta(s-s')\,.
\ee
In the scattering region of $\chi\rightarrow -\infty$, the wavefunction behaves as
\be
\psi_{q,s}\rightarrow \frac{2^{4si}\,\Gamma(-4is)}{\sqrt{\pi}\,|\Gamma(-4is)|}\frac{\Gamma(\mbox{$\frac{1}{2}$}+q+2is)}{|\Gamma(\mbox{$\frac{1}{2}$}+q+2is)|}\,\big(\,e^{2is\chi}+R_q(s) e^{-2is\chi}\,\big)\,,
\ee
where the reflection amplitude is given by $R_q(s)=\frac{\Gamma(\,4is\,)}{\Gamma(\ng-\ng 4is)}\frac{\Gamma(\frac{1}{2}+q-2is)}{\Gamma(\frac{1}{2}+q+2is)}2^{-8si}$. On the other hand, in the forbidden region of $\chi \rightarrow \infty$, the wave function decays again doubly-exponentially as
\be
\psi_{q,s}\rightarrow  N_{q,s} \,
\frac{\sqrt{\pi}\phantom{ai}}{(4e^{\chi})^{q+\frac{1}{2}}}\, e^{-2 e^{\chi}}\,.
\ee
With the prescribed regularization in the above, one finds
\be
Z_{q}(\mbox{$\frac{\beta}{2}$},\mbox{$\frac{\beta}{2}$}) \propto %{\mbox{lim}}
\lim_{\chi_q^c \rightarrow \infty} W_q (\chi_q^c)\, w_q \ng \int_{0}^\infty \ng \ng ds \, \rho_q(s)%s \sinh 2\pi s
\, e^{-\beta \frac{s^2}{2\C}}
\ee
with 
\begin{align}    \label{}
 \rho_q(s)= \frac{\pi \,|\Gamma(\mbox{$\frac{1}{2}$}\ng+\ng q\ng+\ng 2is)|^2}{|\Gamma(\mbox{$\frac{1}{2}$}+q)\,\Gamma(\mbox{$\frac{1}{2}$}\ng+\ng 2is)|^2}\, s \sinh 2\pi s\,, \ \ \ \ W_q (\chi)=\frac{8\,\Gamma(\mbox{$\frac{1}{2}$}\ng+\ng q)^2\, e^{-4 e^{\chi_c}} }{ \pi^2 \, (4e^\chi)^{2q+1}\,w_q}\,.
%\tilde{J}_{1} &\cong  p_{\tau_r}+ p_{\tau_l}+J^m_1\cong 0\,,   \nonumber \\
%\tilde{J}_{2} &\cong  p_{\tau_r}- p_{\tau_l}+J^m_2\cong 0\,,  \nonumber  \\
%\tilde{J}_{3} &\cong  p_{\chi_r}- p_{\chi_l}+J^m_3\cong 0\,, % \nonumber 
\end{align}
Of course, at this point, one may freely adjust the redundant factor $w_q$. % freely. 
Since 
$W_q (\chi^c_q)$ is independent of $s$, we may drop this  in the limit where $\chi^c_q$ goes to infinity.  Hence the leading-order contribution of the two-sided function reads
 \be
Z^{(0)}_{q}(\mbox{$\frac{\beta}{2}$},\mbox{$\frac{\beta}{2}$})= w_q \ng \int_{0}^\infty \ng \ng ds \, \rho_q(s)%s \sinh 2\pi s
\, e^{-\beta \frac{s^2}{2\C}}\,,
\ee
where $w_q$ is not determined at the moment. 

Without insertion of any matter operators  ($q=0$ and $w_0=1$), the above expression agrees with the pure JT result
in (\ref{JTpartition}), {\it i.e.} $Z^{(0)}_{q=0}(\mbox{$\frac{\beta}{2}$},\mbox{$\frac{\beta}{2}$})=Z(\beta)$; This also gives the partition function even in the presence of matter as was argued in  \cite{Penington:2023dql}.  %with $\beta_r+\beta_l=\beta$. 
For $q\neq 0$, the above involves an initial (or final) insertion of boundary matter operators  leading to (two-sided) $2$-point correlation functions in general.

With $q=1$, for instance, one has 
\be
\rho_1= (1+16 s^2)\, s \sinh 2\pi s
\ee
which leads to
 \be
Z^{(0)}_{q=1}(\mbox{$\frac{\beta}{2}$},\mbox{$\frac{\beta}{2}$})= w_1 
Z(\beta) %Z^{(0)}_{q=0}(\mbox{$\frac{\beta}{2}$},\mbox{$\frac{\beta}{2}$})
\Big(
1+ 48\frac{\C}{\beta}+64\pi^2\frac{\C^2}{\beta^2}
\Big)\,.
\ee
Similarly, for the larger value of $q$, one may work out the zeroth-order contribution to the partition function.
Adding the contribution from the first-order perturbation, one has
\be 
Z_q(\mbox{$\frac{\beta}{2}$},\mbox{$\frac{\beta}{2}$})=Z^{(0)}_{q}(\mbox{$\frac{\beta}{2}$},\mbox{$\frac{\beta}{2}$})\, e^{-\frac{\beta}{8\C} \langle \vec{k}|({J^{ m}_3})^2|\vec{k}\rangle}
\ee
where  $\langle \vec{k}|({J^{ m}_3})^2|\vec{k}\rangle=\frac{1}{2}\sum^\infty_{n=1}\big[n(n+1)k_n k_{n+1}+n^2k_n 
\big]$.
Considering now a general matter initial state $\sum_{\vec{k}} c_{\vec{k}} |\vec{k}\rangle$, we need to fix the relative factor $w_q$. There seems no general principle to fix this relative 
factor %especially 
because the $\chi\rightarrow \infty$ limit around the initial and final regularized surfaces is not well understood.   Here we propose to set  the relative factor $w_q=1$ %set to be unity.
and to adjust $\chi^c_{\vec{k}}$ such that $W_q(\chi_{\vec{k}}^c)= W_0(\chi_{\vec{0}}^c)$ in the 
$\chi_{\vec{k}}^{c}\rightarrow \infty$ limit. We then drop the overall factor $ W_0(\chi_{\vec{0}}^c)$ uniformly for any $|\vec{k}\rangle$.  Then for the matter initial state $\sum_{\vec{k}} c_{\vec{k}} |\vec{k}\rangle$,
the zeroth-order two-sided function becomes
\be
Z_{(0)}(\mbox{$\frac{\beta}{2}$},\mbox{$\frac{\beta}{2}$})=\sum_{\vec{k}} |c_{\vec{k}}|^2 \int_{0}^\infty \ng \ng ds \, \rho_{q_{\vec{k}}}(s)%s \sinh 2\pi s
\, e^{-\beta \frac{s^2}{2\C}}
\ee
where $q_{\vec{k}}$ denotes $\sum^\infty_{n=1}\ng  n k_n$. With the above prescription,  
one   has 
%the density of states for each $\vec{k}$ approach that of the pure JT theory
$\rho_{q_{\vec{k}}}(s)\rightarrow \rho_{0}(s)$ in the $s\rightarrow 0$ limit; This also corresponds 
to fixing each density of states to that of the pure JT theory in the zero temperature limit.

\begin{figure}[htbp]   
\begin{center}
\vskip0.1cm
\begin{tikzpicture}[scale=1]
%\draw[blue, thick,-<](-2,0)--++(0,2);\draw[blue,thick] (-2,2)node[left]{\large$t_l$}--++(0,2);
%\draw[thick,red, ->](2,0)--++(0,2);\draw[thick,red] (2,2)node[right]{\large$ t_r$}--++(0,2);
%\draw[blue,decoration={coil,segment length=0.5mm,amplitude=0.15mm},decorate] (-2,2) arc (180:270:2);
\draw %[red,decoration={coil,segment length=0.5mm,amplitude=0.15mm},decorate] 
(2,2) arc (0:360:2); \draw[%red,
decoration={coil,segment length=6mm,amplitude=0.3mm},decorate] (-0.2,3.83) arc (98:262:1.83);
\draw[%red,
decoration={zigzag,
segment length=6mm,amplitude=0.4mm},decorate] (0.1,3.83) arc (86:-86:1.83);
\draw[red] (-0.2,3.83)%node[below,xshift=0.1cm]{$\chi_c$} 
arc (180:360:0.15);
\draw[red] (-0.2,0.21) arc (180:-5:0.15);
%\draw%[thick] 
%(-0.6,0.29) .. controls (-0.0,0.58).. (0.6,0.26);
%\draw[fill=black] (0,0) circle (0.05cm);
%\draw[decoration={zigzag},decorate](-2,0)--(2,0); 
% \draw[thick,dotted](-2,2)--++(4,0);
%\draw[decoration={zigzag},decorate](-2,4)--++(4,0); 
%\draw[decoration={zigzag,segment length=1mm,amplitude=0.5mm},decorate](-2,4)--++(0.2,+0.35); 
%\draw[decoration={zigzag,segment length=1mm,amplitude=0.5mm},decorate](2,4)--++(-0.2,+0.35); 
%\draw[decoration={zigzag,segment length=1mm,amplitude=0.5mm},decorate](-2,0)--++(0.2,-0.35); 
%\draw[decoration={zigzag,segment length=1mm,amplitude=0.5mm},decorate](2,0)--++(-0.2,-0.35); 
%\draw[thick,brown,dotted] (-2,0)--++(4,4);
%\draw[thick,brown,dotted] (+2,0)--++(-4,4);
%\draw (-2,0)--++(4,4);
%\draw (+2,0)--++(-4,4);
\draw[fill,fill opacity=0.02]
%[fill=%fill=yellow!80!black,draw opacity=0.1]  
%[red,decoration={coil,segment length=0.5mm,amplitude=0.15mm},decorate] 
(8,2) arc (0:360:2); \draw[%red,
%fill,fill opacity=0.03,
decoration={coil,segment length=6mm,amplitude=0.3mm},decorate] (6.55,3.7) arc (74:286:1.80);
\draw[%red,
%fill,fill opacity=0.03,
decoration={zigzag,
segment length=5mm,amplitude=0.4mm},decorate] (6.85,3.58) arc (64:-65:1.75);
\draw[red%,fill=black,fill opacity=0.3
] (6.55,3.7)%node[below,xshift=0.1cm]{$\chi_c$} 
arc (160:338:0.16)%;
%\draw[red]
 (6.55,0.24) arc (200:45:0.17);
\draw[fill=black] (6.7,3.65) circle (0.07cm);
\draw[fill=black] (6.71,0.31) circle (0.07cm);
\draw%[red] 
(4.55,1.9) node%[right]
{\small $\beta_l$};\draw%[red] 
(7.55,1.9) node%[right]
{\small $\beta_r$};
\draw%[red] 
(-1.46,1.9) node%[right]
{\small $\beta_l$};\draw%[red] 
(1.51,1.9) node%[right]
{\small $\beta_r$};
\draw[blue] 
(-1.81,2.2) node%[right]
{\tiny $\times$};
\draw[blue] 
(-1.3,3.2) node%[right]
{\tiny $\times$};
\draw[blue] 
(-1.4,0.9) node%[right]
{\tiny $\times$};
%\draw[blue] 
%(1.68,2.68) node%[right]
%{\tiny $\times$};
\draw[blue] 
(1.8,1.6) node%[right]
{\tiny $\times$};
\draw[blue] 
(1.29,3.3) node%[right]
{\tiny $\times$};
\draw[blue] 
(1.21,0.6) node%[right]
{\tiny $\times$};
\draw[blue] 
(4.29,2.2) node%[right]
{\tiny $\times$};
\draw[blue] 
(4.75,3.2) node%[right]
{\tiny $\times$};
\draw[blue] 
(4.6,0.9) node%[right]
{\tiny $\times$};
%\draw[blue] 
%(7.7,2.68) node%[right]
%{\tiny $\times$};
\draw[blue] 
(7.75,1.6) node%[right]
{\tiny $\times$};
\draw[blue] 
(7.35,3.25) node%[right]
{\tiny $\times$};
\draw[blue] 
(7.19,0.6) node%[right]
{\tiny $\times$};
%\draw[blue,decoration={coil,segment length=0.5mm,amplitude=0.15mm},decorate] (6,4) arc (90:270:2);\draw[red,decoration={coil,segment length=0.5mm,amplitude=0.15mm},decorate] (6,4) arc (90:-90:2); \draw[fill=black] (6,0) circle (0.05cm);
 %\draw[thick,dotted](4,2)--++(4,0);\draw[fill=black] (6,0) circle (0.05cm);
%\draw[decoration={zigzag},decorate](-2,4)--++(4,0); 
%\draw[blue,decoration={coil,segment length=0.5mm,amplitude=0.2mm},decorate] (4,2) arc (180:90:2);
%\draw[red,decoration={coil,segment length=0.5mm,amplitude=0.2mm},decorate] (8,2) arc (0:90:2); 
%\draw[fill=black] (6,4) circle (0.05cm);
\draw[thick,dotted](9.7,1.8)--++(2.5,0)node[right]{\small $\hat{\varphi}_{r/l}$};
\draw [line width=1pt, double distance=1.7pt,
             arrows = {-%Stealth[length=0pt 1.5 0]}] 
Classical TikZ Rightarrow[length=0pt 1 0]}] 
(10.95,0.75)node[below]{\small $k$} --++ (0,0.95);
\draw [line width=1pt, double distance=1.7pt,
             arrows = {-%Stealth[length=0pt 1.5 0]}] 
Classical TikZ Rightarrow[length=0pt 1 0]}] 
(10.42,1.9)--++ (0,0.95)node[above]{\small $k\ng-\ng1\,\,\,$};
\draw [line width=1pt, double distance=1.7pt,
             arrows = {-%Stealth[length=0pt 1.5 0]}] 
Classical TikZ Rightarrow[length=0pt 1 0]}] 
(11.45,1.9) --++ (0,0.95) node[above]{\small $\,\,\, k\ng +\ng1$};
\draw%[red] 
(0.15,-0.5) node%[right]
{\small $(a)$};\draw%[red] 
(6.2,-0.5) node%[right]
{\small $(b)$};
\draw%[red] 
(10.95,-0.5) node%[right]
{\small $(c)$};
\end{tikzpicture}
\vskip-0.1cm
\caption{\small In (a) and (b), we draw the two-side evolution for  JT theory with matter operators inserted along the left right evolutions. In these diagrams, the matter insertions are marked by cross symbols.
 (a) depicts the case where one starts from and ends with the matter vacuum state.
% $|\vec{0}\rangle$. 
(b) describes the case with general matter initial  state. 
With a left or right insertion of matter operator,
%the both involve nontrivial matter excitations along the evolution. 
an  $N_m=k$  state %belonging to 
before the insertion turns into a linear combination of
 a $k+1$ and a $k-1$ state, which is depicted in (c).
%Correlation functions defined 
%in this way are dressed with gravitational fluctuations. 
}
\label{Fig3}
\end{center}
\end{figure}
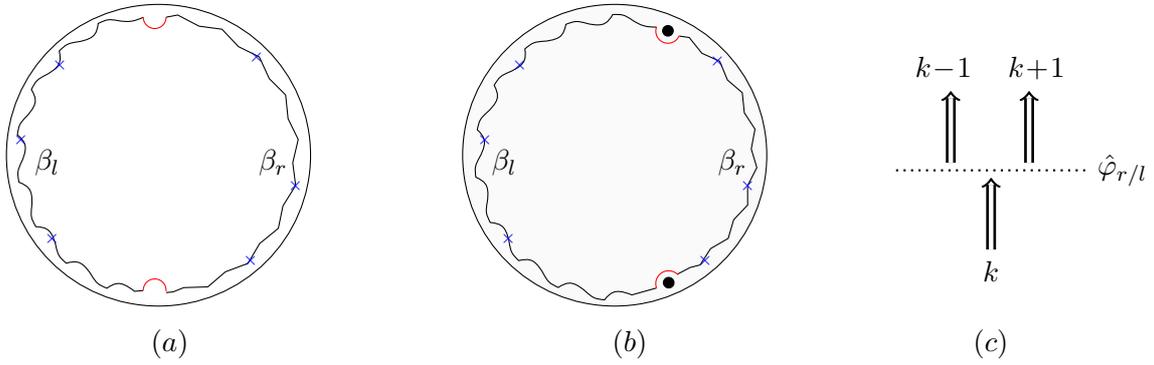 
% %

In general, the evolution by $H_{l/r}$ preserves 
$N_m$ and $C_m$ quantum numbers as they commute with the left and right Hamiltonians.
%Then to all orders 
Thus, for instance,  one may start from a matter initial state which belongs to 
a particular matter sector specified by the eigenvalues of $N_m$ and $C_m$.  Then,
along the evolution by $H_{l/r}$, matter states  stay within the initially prescribed sector. 

As depicted in Figures \ref{Fig3}{\color{blue}a} and \ref{Fig3}{\color{blue}b}, 
%But, even with this kind of initial states, 
one may consider inserting matter operators along the boundary trajectories.
%, one finds  mixing  between sectors occurs in a rather complicated manner
% as depicted in Figure \ref{Fig3}. 
% depicted in the left side of Figure \ref{Fig3}, 
Let us introduce the corresponding correlation function defined  by
\be
G^E_{\,I}({y^r_1, \cdots, y^r_{n_r}; y^l_1, \cdots, y^l_{n_l}})=\langle  %{\cal T}_r
\ng 
\prod^{n_r}_{k=1} \hat{\varphi}_{r}(-i y^r_{k}) \,%{\cal T}_l\ng
\prod^{n_l}_{k'=1} \hat{\varphi}_{l}(-i y^l_{k'})\rangle_I\,,
%\rangle \ \ \ (k_{r/l}=1,2, \cdots, n_{r/l})
\ee
with $y^{r/l}_{\,k} \in [0,\beta_{r/l}]$. In this case, as a pair of the
% left and right depending on the choice of slicing
Euclidean times $y_l$ and $y_r$ evolves\footnote{The two-sided evolution here is very much dependent upon ways of slicing $y_l$ and $y_r$.  However the final answer should be independent of slicing, as dictated by any gravity theories.}\ng, one will 
encounter insertions of  left or right matter operators order by order. Now in between each successive
encounters, let us focus on a  state belonging to an $N_m=k$ sector with $k\ge 0$, where %whose 
its evolution remains within the sector in between the encounters. 
This 
%component of 
state will eventually encounter a left or right operator $\hat\varphi_{l/r}$ at a certain
slice which is denoted by the dotted line in Figure \ref{Fig3}{\color{blue}c}. 
%Right after the 
%encounter,  the $N_m=k$(-sector) state with  $k \ge 1$ turns into a linear combination of $N_m=k-1$ and $k+1$
%states while the $N_m=0$ state (with $k=0$)  becomes simply an $N_m=1$ state. 
Right after the 
encounter, the $N_m=k$ state turns into a linear combination of $N_m\ng=\ng k-1$ and $k+1$
states where %, of course, 
 the $N_m\ng =\ng -1$ state (with $k\ng=\ng 0$) does not exist and  should be removed additionally.
This process goes on with next encounters of operators. 
It then follows that %, consequently, %It is then rather straightforward to check that 
any  correlation function, defined with $|\Phi_I\rangle=|\Phi_F\rangle$ that belongs to a particular $N_m=k$ sector, %with $k\ge 0$, 
vanishes if $n_r+n_l$ is odd.  
 
%In particular 
%To illustrate the above explicitly, 
For the sake of %explicit 
illustration,
let us 
 consider the case %with $|\Phi_I\rangle=|\Phi_F\rangle$
where one  starts from and ends with the matter vacuum state. Of course,
 for the partition function, 
the corresponding evolution stays %occurs 
within the matter vacuum sector. 
%But, even in this case, 
Now once we add matter operators along the %left and right
 evolution,
the state  no longer stays within the matter vacuum sector  and mixing between sectors will occur as described in the above.   
The first nontrivial example is the two-point correlation function  $G^E_{|\vec{0}\rangle}({y^r_1;y^l_1})$ where we further assume $\beta_r=\beta_l =\beta/2$ and $y^r_1 < y_1^l$ for simplicity.
%and let us take $y^$
Then one may evolve the system with $H_{tot}$ with Euclidean time $0<y<{\beta}/{2}$. For
$0<y< y^r_1$, the system remains within the matter vacuum sector. Then, for  
$y^r_1 < y < y^l_1$, the state belongs to the $N_m=1$ %and $C_m=0$ 
sector.  
Finally for $y^l_1 < y < \beta/2$, the evolution is restricted to the matter vacuum sector due to the final state condition.
%vanishes since $ \langle \vec{0}|\,\hat{\varphi}_{r/l}\, |\vec{0}\rangle=0$.
An explicit evaluation of this two-point function 
does not seem to be so straightforward. 
%Also we are not sure 
Neither is it clear to us % It is not clear to us %either 
how the above correlation functions are related to the 
%gravity-dressed 
conventional
correlation functions 
%in various approaches 
in literatures
\cite{Mertens:2017mtv,Blommaert:2018oro,Kitaev:2018wpr,Yang:2018gdb,Iliesiu:2019xuh,Lin:2022rbf,Mertens:2022irh}. %(see also references therein).
%These  require 
Further studies are required in this direction.

%%%%%%%%%%%%%%%%%%%%%%%%
\section{Conclusion}
%%%%%%%%%%%%%%%%%%%%%%%%%%%%%%%%%%%
%Two-dimensional JT gravity with negative cosmological constant has extensively been studied as a concrete model for the AdS/CFT correspondence. Because of the two-dimensional nature of gravity, its dynamics is described by Schwarzian action defined along cut-off boundary trajectories, and  its structure is simple enough to allow canonical quantization in Lorentzian signature. Though the Hilbert space of Schwarzian theory is not a tensor product of one-sided Hilbert spaces, the two-sided Hilbert space is well-defined and the time evolution of a state can be described by the total Hamiltonian $H_{r}+H_{l}$. The left and right boundary algebras, ${\cal A}_{l/r}$, including $H_{l/r}$ as an element can be constructed and satisfy $[{\cal A}_{l}, {\cal A}_{r}]=0$.  

 %In pure JT gravity, $H_{r}- H_{l}=0$ and so the time evolution given by $H_{r}-H_{L}$ has no meaning. On the other hand, in JT gravity coupled to matter, the  additional time evolution given by $H_{r}- H_{l}$ becomes also physical,  which can be interpreted  that the time evolutions  along the left and right cut-off trajectories are physically independent.  
%In JT gravity coupled to bulk matter, the boundary algebras are shown to be a von Neumann factor of type II$_{\infty}$   
%and  the entanglement entropy in Lorentzian signature is matched to the result by the Euclidean path-integral~
%\cite{Penington:2023dql}.  

In this paper, we have presented the detailed canonical quantization of JT gravity coupled to a massless scalar field. Especially, we have identified the bulk matter charges $J^{m}_{i}$ explicitly,
and shown that the (matter) number operator $N_{m}$ and the Casimir $C_m$ commute with the boundary Hamiltonians $H_{l/r}$.  This allows us to choose the simultaneous eigenstate of $H_{l}$, $H_{r}$, $N_{m}$ and $C_m$ in the two-sided Hilbert space.  And then we  computed some simultaneous eigenfunctions  in the two-sided Hilbert space. %and partition function. 
In the pure JT gravity, we reproduced the well-known eigenfunction given by a modified Bessel function. In \cite{Lin:2022zxd}, the two-sided version of the disk partition function was 
proposed by starting two-sided boundary evolution from  an initial geodesic curve connecting two slightly separated boundary points in the bottom region of the disk and ending up with  a final geodesic curve between again  two slightly separated boundary points in the top region. 
From this definition of the two-sided partition function, the disk partition function was reproduced
in the same reference.
In the presence of matter, one may additionally arrange  initial and final states including the matter part  before and after the initial and final regularized curves, by which the bulk of the disk
is affected in general. Thus we have introduced
a two-sided correlation function in the presence of prescribed matter states. In particular, we %proposed the way
 tried to specify the   prescribed
 states at the initial and final regularized 
curves generalizing the proposal of \cite{Lin:2022zxd}. 
In JT gravity with a massless scalar, the eigenfunction of $H_{tot}$ 
is shown to be given by a Wittaker function and the two-sided correlation function for $\beta_{r}=\beta_{l}= \beta/2$  is evaluated perturbatively for some simple initial states.
% such that  appropriate initial and final matter states are arranged. 
One may additionally  insert boundary matter operators along the two-sided evolution 
leading to the higher two-sided  correlation functions. We have investigated some basic properties of these two-sided correlation functions. % in this note.

 %and the partition function is matched to the well-known expression of the one-sided partition function.   We have also introduced  two-sided correlation functions and investigated their basic properties. 

%Many questions remain unanswered.  The physical implication of the two-sided partition function in the presence of matter is not entirely clear to us. This two-sided partition function involves  two Euclidean evolution parameters $\beta_{r}$ and $\beta_{l}$ depending on the boundary locations of the initial and final states. Although these parameters seem natural from the viewpoint of the two-sided evolution,  its precise physical meaning has to be clarified. It would also be interesting to clarify the regularization procedure of the initial and final curves more clearly. In addition, the properties and physical implications of the two-sided correlation function require further studies.

The two-sided correlation functions we have introduced require the specific
regularization procedure of initial (or final) state at the initial (or final) curve, which is not so 
well-motivated unfortunately. Instead one may provide some controlled initial (or final) state there and the resulting two-sided correlation functions may be directly related to the conventional correlation functions in \cite{Mertens:2017mtv,Blommaert:2018oro,Kitaev:2018wpr,Yang:2018gdb,Iliesiu:2019xuh,Lin:2022rbf,Mertens:2022irh}.     However the precise guiding principle to construct such initial (or final) state is lacking at this stage. 
%However the precise
%guiding principle to construct such initial (or final) state is not clear to us.} 
Further investigations are required in this direction.
%and check our proposal in the computation of the partition function in the presence of matter.
%As alluded before, the two-sided partition function  in JT gravity coupled to matter should have generically  two temperature parameters $\beta_{r}$ and $\beta_{l}$. These parameters seem natural from the two-sided picture of the system. %However, it is not so clear how to interpret
%interpretation of 
%Further investigations of this two-sided partition function are required %in terms of the dual boundary theory 
%especially when 
%$\beta_r\neq \beta_l$.
% It would also be interesting to clarify the regularization procedure more clearly and check our proposal in the computation of the partition function in the presence of matter.
 %In addition, the properties and  physical implications of 
%two-sided correlation functions are poorly understood and  require further studies.
% how to understand these parameters. %, especially when $\beta_{1}+\beta_{2} \neq 2\pi$. 
%It would be quite interesting to explore further the meaning of these parameters. 
In addition we have not considered the bulk wormhole contribution~\cite{Penington:2023dql}. It would be interesting to include its effect at the level of  the partition function and to consider the factorization issues.
%when the bulk matter exist. 

%Hence in the $n$-point functions built out of 
%\section{Explicit construction with $m^2=0$}\label{sec3}
\subsection*{Acknowledgement}
%We thank Juan Maldacena for enlightening discussions for 
%clarifying the nature of interactions in traversable wormhole systems.
We would like to thank Andreas Gustavsson for careful reading of the manuscript.
CK thanks CCNY for hospitality where part of this work was done.
DB was
supported in part by
NRF Grant RS-2023-00208011 %2020R1A2B5B01001473, 
and by  Basic Science Research Program
through NRF %National Research Foundation 
funded by the Ministry of Education
(2018R1A6A1A06024977). 
 C.K.\ was supported by NRF Grant 2022R1F1A1074051.
S.-H.Y. was supported by 
%the National Research Foundation of Korea(NRF) 
NRF Grant  %with the grant number NRF- 
2021R1A2C1003644 and supported  by Basic Science Research Program through the NRF funded by the Ministry of Education (NRF-2020R1A6A1A03047877).

\appendix

%%%%%%%%%%%%%%%%%%%%%%%%%%
\section{Representations of the matter charges }\label{AppA}
%%%%%%%%%%%%%%%%%%%%%%%%%%
%%%%%%%%%%%%%%%
\renewcommand{\theequation}{A.\arabic{equation}}
  \setcounter{equation}{0}
 %%%%%%%%%%%%%%%%%%%%%%%%%%
In this appendix, we show that the matter Hilbert space $\mathcal{H}_m$
is decomposed of negative discrete series of irreducible representations 
$\mathcal{D}_j^-$ of SL(2,{\bf R}) (see~\cite{Kitaev:2017hnr} for a review of SL(2,{\bf R}) representation). Introducing
\begin{equation}
J^m_\pm = J^m_2 \pm i J^m_3,
\end{equation}
we can rewrite the SL(2,{\,\bf R}) algebra as
\begin{equation}
[J^m_1, J^m_\pm] = \pm J^m_\pm, \qquad
[J^m_+, J^m_-] = -2J^m_1.
\end{equation}
Thus $J^m_\pm$ may be considered as raising/lowering operators for the 
eigenstates of $J_1^m$. The Casimir operator $C_m$ becomes
\begin{equation}
C_m = -(J_1^m)^2 - J_1^m + J_-^m J_+^m\,.
\end{equation}
In terms of the operators $a$ and $a^\dagger$ in \eqref{scalarf},
$J_\pm^m$ are given by
\begin{align}
{J}^m_+ &=\sum^\infty_{n=1}\sqrt{n(n+1)}a^\dagger_n a_{n+1}\,,  \nonumber  \\
{J}^m_- &=\sum^\infty_{n=1}\sqrt{n(n+1)}a_n a^\dagger_{n+1}\,, % \nonumber 
\end{align}
Note that, for each pair of adjacent oscillators, 
$J_\pm^m$ shift the oscillator number by one to the left/right, respectively.
In particular, $J_+^m$ annihilates states 
$|k\rangle \equiv |k000\cdots\rangle$ for any $k$,
\begin{equation}
J_+^m |k\rangle = 0.
\end{equation}
Since 
\begin{equation}
J_1^m |k\rangle = -k |k\rangle \,, \ \ \ 
C_m |k\rangle = k(1-k)|k\rangle \,, \ \ \  N_m |k\rangle = k|k\rangle\,,
\end{equation}
we can identify $|k\rangle $ as the highest weight state of the 
representation $\mathcal{D}_k^-$ (with $N_m=k$) of SL(2,{\bf R}) in negative discrete series.
Then applying $J_-^m$, we obtain basis vectors of the representation
$\{ | l \rangle\}$ with $l = k, k+1, \ldots$ which are eigenstates of $J_1^m$
with $ J_1^m |l\rangle = -l |l\rangle $. Normalized vectors are
\begin{equation}
|l \rangle \equiv 
       \mbox{$ \sqrt{ \frac{(2k-1)!}{(k+l-1)!\,(l-k)!}}\, (J_-^m)^{l-k}$}  |k \rangle \,.
\end{equation}

Recall that the number operator $N_m$ commutes with $J_i^m$'s. 
Then, from \eqref{nj1}, we see that $|l \rangle$ consists of the oscillator 
states $| \vec{k} \rangle $ with
\begin{equation}
\sum_{n=1}^{l-k+1} k_n = k\,, \qquad
\sum_{n=1}^{l-k+1} n k_n = l\,.
\end{equation}
Note that the upper limit of the summation range is limited by $l-k+1$.
As $l$ increases, there are more oscillator states 
involved to make a particular $| l \rangle$ state. For instance,
for $l=k+1$ and $k+2$, we get
\begin{align}
|k+1 \rangle &=\mbox{$ \frac1{\sqrt{2k}}$} J_-^m |k \rangle
              = |k-1,1,0,0,\cdots\rangle \,, \nonumber\\
|k+2 \rangle &= \mbox{$\frac1{\sqrt{4k(2k+1)}} $}(J_-^m)^2 |k \rangle \nonumber\\
 &= \mbox{$\sqrt{\frac{2(k-1)}{2k+1}}$}\,|k-2,2,0,0,\cdots \rangle
   +\mbox{$ \sqrt{\frac{3}{2k+1}}$}\,|k-1,0,1,0,\cdots \rangle \,.
\end{align}
In this example, applying $J_+^m$ to two oscillator states
in $|k+2 \rangle$ would result in the same $|k-1,1,0,0,\cdots \rangle$ which
is nothing but $|k+1 \rangle$. This implies that the orthogonal combination
\begin{equation}
|\mbox{$\widetilde{k\ng+\ng2}$}\rangle =\mbox{$ \sqrt{\frac{3}{2k+1}}$}\,|k-2,2,0,0,\cdots \rangle
   - \mbox{$\sqrt{\frac{2(k-1)}{2k+1}}$}\,|k-1,0,1,0,\cdots \rangle
\end{equation}
with $k \ge 2$
should be annihilated by $J_+^m$, which can easily be checked. 
Then, we see that $|\widetilde{k\ng+\ng 2} \rangle$ is the highest weight state of 
a new irreducible representation $\mathcal{D}_{k+2}^-$ (with $N_m=k$) which is obtained by
applying $J_-^m$ successively to $|\widetilde{k\ng +\ng 2} \rangle$.

It is clear to generalize this procedure. If applying $J_-^m$ increases
the number of oscillator states which participate in the linear combination,
one would get highest weight states of new irreducible representations by
considering the orthogonal linear combinations of the states. In this way,
the matter Hilbert space $\mathcal{H}_m$ can be decomposed of negative
discrete series of irreducible representations 
$\mathcal{D}_j^-$ of SL(2,{\bf R}).

 %%%%%%%%%%%%%%%%%%%%%%%%%%

%%%%%%%%%%%%%%%%%%%%%%%%%%
\section{Gauge-fixing}\label{AppB}
%%%%%%%%%%%%%%%%%%%%%%%%%%
%%%%%%%%%%%%%%%
\renewcommand{\theequation}{B.\arabic{equation}}
  \setcounter{equation}{0}
 %%%%%%%%%%%%%%%%%%%%%%%%%%
In this appendix, we present some details on the gauge-fixing procedure. Though we use the commutator notation of quantum mechanics, it may be understood as the corresponding Poisson bracket in the context of classical mechanics. Note that, in the classical setup, the last terms in $J^{r/l}_{2}$ and $J^{r/l}_{3}$ %expressions 
in~\eqref{rlcharges} do not appear. As  mentioned in Section \ref{sec2}, the condition
$|\tau_r-\tau_l| <\pi$ will be assumed.
%First of all, let us note that 
Let us begin with
\begin{equation} \label{}
i\big[\tilde{J}_{1}, \,\textstyle{\frac{1}{2}}(\tau_{r}\ng + \ng \tau_{l}) \big] = 1\,, 
\end{equation}
which allows us to fix the gauge, $\tau_{r}+\tau_{l}=0$. Upon %Under 
this gauge choice, we may see that
\begin{equation} \label{}
i\big[\tilde{J}_{2}, \,\textstyle{\frac{1}{2}}(\tau_{r}\ng -\ng \tau_{l}) \big] =  i\big[\tilde{J}_{2}, \, \pm\tau_{r/l}  \big] = \cos\tau_{r}=\cos\tau_{l}\,.
\end{equation}
%
%In pure JT gravity, this commutator relation can be rewritten, by using the cut-off trajectory %expression in~\eqref{cutofftra}  as
%
%\begin{equation} \label{}
%i\Big[\tilde{J}_{2},~ u  \Big] = 1\,,
%\end{equation}
%
%which allows us to set $u=0$. Then, this leads to the gauge constraint $\tau_{r}=\tau_{l}=0$ from the condition $\tau_{r/l}(u=0)=0$. 
Thus we may set $\tau_r=\tau_l=0$ where we used the condition $|\tau_r-\tau_l| <\pi$.
Now, with $\tau_r=\tau_l=0$, we find
\begin{equation} \label{}
i\big[\tilde{J}_{3}, \, \pm\chi_{r/l}  \big] = 1 \,, 
\end{equation}
which allows us to fix the gauge $\chi_r-\chi_l=0$. This completes our gauge-fixing procedure.
Classically, starting with the relevant solutions in \cite{Bak:2021qbo}, one may work out the
corresponding gauge transformations explicitly which lead to the fully gauge-fixed forms of solutions.
%Finally, by using an appropriate rotation generated by $\tilde{J}_{3}$ acting on $e^{\chi_{r/l}}= %A_{r/l} \mp C_{r/l}$, one can show that $e^{\chi_{r}}=e^{\chi_{l}}$. 
 
%%%%%%%%%%%%%%%%%%%%%%%%%%
\section{Classical bulk solutions}\label{AppC}
%%%%%%%%%%%%%%%%%%%%%%%%%%
\renewcommand{\theequation}{C.\arabic{equation}}
  \setcounter{equation}{0}
Here we summarize classical bulk solutions of JT gravity. See~\cite{Bak:2021qbo} for more details.  
 Under the vanishing boundary condition,  the scalar equation (\ref{massless}) with $m=0$ 
is solved by
 %bulk scalar field solution is given by 
 %
\begin{equation} \label{}
\varphi = \sum_{n=1}^{\infty}{\bf a}_{n} \sin n (\mu+\frac{\pi}{2}) \cos n(\tau-\tau_{n}).
\end{equation}
In the main text, we introduced complex coefficients $a_{n}$'s by the relations %which are rescaled  from ${\bf a}_{n}$'s as 
\begin{equation} \label{}
a_{n} \equiv \frac{\sqrt{n\pi}}{2}e^{in\tau_{n}}\, {\bf a}_{n}  \,, \qquad a^{\dagger}_{n} \equiv \frac{\sqrt{n\pi}}{2} e^{-in\tau_{n}}\, {\bf a}_{n} \,.  
\end{equation}
Then, the above %scalar field 
solution   can be rewritten as \eqref{scalarf}.

Now, let us return to the dilaton field $\phi$.  As was shown in~\cite{Bak:2021qbo}, the classical solution of the dilaton $\phi$ is obtained in the form of
\begin{equation} \label{}
\phi = \bar{\phi}L\frac{\cos\tau}{\cos\mu} + \sum_{m,n=1}^{\infty}{\bf a}_{m}{\bf a}_{n}\phi_{m,n}\,, 
\end{equation}
where the explicit expressions of $\phi_{m,n}$'s are given by
\begin{align}    \label{}
\phi_{n,n}\,\, &= {\textstyle \frac{(-1)^{n}n}{8(4n^{2}-1)} }(2n\cos2n\mu +\sin 2n\mu +\tan\mu) \cos 2n(\tau-\tau_{n}) -\frac{n^{2}}{4}( 1+\mu\tan\mu) \nonumber \\
\phi_{n,n\ng+\ng 1} &= \phi_{n+1,n} \nonumber \\
& = {\textstyle\frac{(-1)^{n+1}}{16(2n+1)\cos\mu}} \big[(n\ng+\ng1)\sin2n\mu \ng+\ng n\sin(n\ng+\ng 2)\mu\big]\ng \cos[(2n\ng+\ng1)\tau\ng-\ng (n\ng+\ng 1)\tau_{n+\ng1}\ng-\ng n\tau_{n})]  \nonumber  \\
& \qquad - {\textstyle \frac{n(n+1)}{8\cos\mu} }(\tan\mu +\mu)\cos[\tau- (n+1)\tau_{n+\ng 1} + n\tau_{n}]\,, \nonumber \\
\phi_{m,n} \,\,&= {\textstyle \frac{mn}{8\cos\mu} }\Big[\cos[n(\tau-\tau_{n}) -m(\tau-\tau_{m})] \Big( \textstyle{ \frac{\sin(n-m+1)(\mu+\frac{\pi}{2})}{(n-m+1)(n-m)} -  \frac{\sin(n-m-1)(\mu+\frac{\pi}{2})}{(n-m-1)(n-m)}  } \Big)  \nonumber  \\
& \qquad +\cos[n(\tau-\tau_{n}) + m(\tau-\tau_{m})] \Big( \textstyle{ \frac{\sin(n+m+1)(\mu+\frac{\pi}{2})}{(n+m+1)(n-m)} -  \frac{\sin(n+m-1)(\mu+\frac{\pi}{2})}{(n+m-1)(n-m)}  } \Big)\Big]\,. \label{bulksol}
\end{align}
The asymptotic behaviors of these solutions as $\mu\rightarrow \mu_{c}^{r/l}$ read as 
\begin{align}    \label{}
\phi_{n,n}\,\, &= -\frac{n^{2}}{4}(1+\mu\tan\mu) + {\cal O}(\cos^{2}\mu)\,,  \nonumber \\
\phi_{n,n+1}&= \phi_{n+1,n} = -\frac{n(n+1)}{8}(\sin\mu\ng+\ng\mu\sec\mu)\cos[\tau\ng-\ng(n\ng+\ng1)\tau_{n\ng+\ng1}\ng+\ng n\tau_{n}]\ng  +\ng {\cal O}(\cos^{2}\mu)\,,\nonumber 
\end{align}
and all the remaining $\phi_{n,m} =  {\cal O}(\cos^{2}\mu)$.  This asymptotic form leads to \eqref{phicutoff}  and the expressions for  $Q^{r/l}_{\, i}$ %expressions 
in~\eqref{Qexp}.

%\end{appendix}


\begin{thebibliography}{99}\label{bibg}


%\cite{Jackiw:1984je}
\bibitem{Jackiw:1984je}
R.~Jackiw,
``Lower Dimensional Gravity,''
Nucl. Phys. B \textbf{252}, 343-356 (1985).
%doi:10.1016/0550-3213(85)90448-1
%728 citations counted in INSPIRE as of 11 Aug 2022

%\cite{Teitelboim:1983ux}
\bibitem{Teitelboim:1983ux}
C.~Teitelboim,
``Gravitation and Hamiltonian Structure in Two Space-Time Dimensions,''
Phys. Lett. B \textbf{126}, 41-45 (1983).
%doi:10.1016/0370-2693(83)90012-6
%707 citations counted in INSPIRE as of 11 Aug 2022


%\cite{Mertens:2022irh}
\bibitem{Mertens:2022irh}
T.~G.~Mertens and G.~J.~Turiaci,
``Solvable Models of Quantum Black Holes: A Review on Jackiw-Teitelboim Gravity,''
[arXiv:2210.10846 [hep-th]].
%14 citations counted in INSPIRE as of 24 Feb 2023


%\cite{Sarosi:2017ykf}
\bibitem{Sarosi:2017ykf}
G.~S\'arosi,
``AdS$_{2}$ holography and the SYK model,''
PoS \textbf{Modave2017}, 001 (2018)
%doi:10.22323/1.323.0001
[arXiv:1711.08482 [hep-th]].
%156 citations counted in INSPIRE as of 11 Aug 2022

%\cite{Trunin:2020vwy}
\bibitem{Trunin:2020vwy}
D.~A.~Trunin,
``Pedagogical introduction to the Sachdev\textendash{}Ye\textendash{}Kitaev model and two-dimensional dilaton gravity,''
Usp. Fiz. Nauk \textbf{191}, no.3, 225-261 (2021).
%doi:10.3367/UFNe.2020.06.038805
%[arXiv:2002.12187 [hep-th]].
%37 citations counted in INSPIRE as of 23 Mar 2023

%\cite{Maldacena:2016upp}
\bibitem{Maldacena:2016upp}
J.~Maldacena, D.~Stanford and Z.~Yang,
``Conformal symmetry and its breaking in two dimensional Nearly Anti-de-Sitter space,''
PTEP \textbf{2016}, no.12, 12C104 (2016)
%doi:10.1093/ptep/ptw124
[arXiv:1606.01857 [hep-th]].
%708 citations counted in INSPIRE as of 21 Aug 2022

%\cite{Jensen:2016pah}
\bibitem{Jensen:2016pah}
K.~Jensen,
``Chaos in AdS$_2$ Holography,''
Phys. Rev. Lett. \textbf{117}, no.11, 111601 (2016)
%doi:10.1103/PhysRevLett.117.111601
[arXiv:1605.06098 [hep-th]].
%508 citations counted in INSPIRE as of 15 Aug 2022

%\cite{Engelsoy:2016xyb}
\bibitem{Engelsoy:2016xyb}
J.~Engels\"oy, T.~G.~Mertens and H.~Verlinde,
``An investigation of AdS$_{2}$ backreaction and holography,''
JHEP \textbf{07}, 139 (2016)
%doi:10.1007/JHEP07(2016)139
[arXiv:1606.03438 [hep-th]].
%422 citations counted in INSPIRE as of 17 Aug 2022



%\cite{Stanford:2017thb}
\bibitem{Stanford:2017thb}
D.~Stanford and E.~Witten,
``Fermionic Localization of the Schwarzian Theory,''
JHEP \textbf{10}, 008 (2017)
%doi:10.1007/JHEP10(2017)008
[arXiv:1703.04612 [hep-th]].
%279 citations counted in INSPIRE as of 17 Feb 2023



%\cite{Harlow:2018tqv}
\bibitem{Harlow:2018tqv}
D.~Harlow and D.~Jafferis,
``The Factorization Problem in Jackiw-Teitelboim Gravity,''
JHEP \textbf{02}, 177 (2020)
%doi:10.1007/JHEP02(2020)177
[arXiv:1804.01081 [hep-th]].
%148 citations counted in INSPIRE as of 12 Feb 2023



%\cite{Penington:2023dql}
\bibitem{Penington:2023dql}
G.~Penington and E.~Witten,
``Algebras and States in JT Gravity,''
[arXiv:2301.07257 [hep-th]].
%0 citations counted in INSPIRE as of 24 Jan 2023





%\cite{Jafferis:2019wkd}
\bibitem{Jafferis:2019wkd}
D.~L.~Jafferis and D.~K.~Kolchmeyer,
``Entanglement Entropy in Jackiw-Teitelboim Gravity,''
[arXiv:1911.10663 [hep-th]].
%28 citations counted in INSPIRE as of 24 Jan 2023


%\cite{Marolf:2012xe}
\bibitem{Marolf:2012xe}
D.~Marolf and A.~C.~Wall,
``Eternal Black Holes and Superselection in AdS/CFT,''
Class. Quant. Grav. \textbf{30}, 025001 (2013)
%doi:10.1088/0264-9381/30/2/025001
[arXiv:1210.3590 [hep-th]].
%73 citations counted in INSPIRE as of 25 Feb 2023





%\cite{Leutheusser:2021qhd}
\bibitem{Leutheusser:2021qhd}
S.~Leutheusser and H.~Liu,
``Causal connectability between quantum systems and the black hole interior in holographic duality,''
[arXiv:2110.05497 [hep-th]].
%42 citations counted in INSPIRE as of 02 Mar 2023
%Copy to Clipboard Download


%\cite{Leutheusser:2021frk}
\bibitem{Leutheusser:2021frk}
S.~Leutheusser and H.~Liu,
``Emergent times in holographic duality,''
[arXiv:2112.12156 [hep-th]].
%43 citations counted in INSPIRE as of 02 Mar 2023


%\cite{Witten:2021unn}
\bibitem{Witten:2021unn}
E.~Witten,
``Gravity and the crossed product,''
JHEP \textbf{10}, 008 (2022)
%doi:10.1007/JHEP10(2022)008
[arXiv:2112.12828 [hep-th]].
%46 citations counted in INSPIRE as of 02 Mar 2023

%\cite{Chandrasekaran:2022cip}
\bibitem{Chandrasekaran:2022cip}
V.~Chandrasekaran, R.~Longo, G.~Penington and E.~Witten,
``An algebra of observables for de Sitter space,''
JHEP \textbf{02}, 082 (2023)
%doi:10.1007/JHEP02(2023)082
[arXiv:2206.10780 [hep-th]].
%49 citations counted in INSPIRE as of 02 Mar 2023


%\cite{Chandrasekaran:2022eqq}
\bibitem{Chandrasekaran:2022eqq}
V.~Chandrasekaran, G.~Penington and E.~Witten,
``Large N algebras and generalized entropy,''
[arXiv:2209.10454 [hep-th]].
%19 citations counted in INSPIRE as of 02 Mar 2023



%\cite{Kolchmeyer:2023gwa}
\bibitem{Kolchmeyer:2023gwa}
D.~K.~Kolchmeyer,
``von Neumann algebras in JT gravity,''
[arXiv:2303.04701 [hep-th]].
%0 citations counted in INSPIRE as of 09 Mar 2023


%\cite{Bak:2021qbo}
\bibitem{Bak:2021qbo}
D.~Bak, C.~Kim, S.~H.~Yi and J.~Yoon,
``Python\textquoteright{}s lunches in Jackiw-Teitelboim gravity with matter,''
JHEP \textbf{04}, 175 (2022)
%doi:10.1007/JHEP04(2022)175
[arXiv:2112.04224 [hep-th]].
%3 citations counted in INSPIRE as of 11 Feb 2023



%\cite{Lin:2022zxd}
\bibitem{Lin:2022zxd}
H.~W.~Lin, J.~Maldacena, L.~Rozenberg and J.~Shan,
``Looking at supersymmetric black holes for a very long time,''
[arXiv:2207.00408 [hep-th]].
%15 citations counted in INSPIRE as of 17 Feb 2023


%\cite{Almheiri:2014cka}
\bibitem{Almheiri:2014cka}
A.~Almheiri and J.~Polchinski,
``Models of AdS$_{2}$ backreaction and holography,''
JHEP \textbf{11}, 014 (2015)
%doi:10.1007/JHEP11(2015)014
[arXiv:1402.6334 [hep-th]].
%459 citations counted in INSPIRE as of 16 Aug 2022


%\cite{Bagrets:2016cdf}
\bibitem{Bagrets:2016cdf}
D.~Bagrets, A.~Altland and A.~Kamenev,
``Sachdev\textendash{}Ye\textendash{}Kitaev model as Liouville quantum mechanics,''
Nucl. Phys. B \textbf{911}, 191-205 (2016).
%doi:10.1016/j.nuclphysb.2016.08.002
%[arXiv:1607.00694 [cond-mat.str-el]].
%235 citations counted in INSPIRE as of 04 Mar 2023


%\cite{Lin:2019qwu}
\bibitem{Lin:2019qwu}
H.~W.~Lin, J.~Maldacena and Y.~Zhao,
``Symmetries Near the Horizon,''
JHEP \textbf{08}, 049 (2019)
%doi:10.1007/JHEP08(2019)049
[arXiv:1904.12820 [hep-th]].
%80 citations counted in INSPIRE as of 09 Feb 2023




%\cite{Marolf:2008hg}
\bibitem{Marolf:2008hg}
D.~Marolf and I.~A.~Morrison,
``Group Averaging for de Sitter free fields,''
Class. Quant. Grav. \textbf{26}, 235003 (2009)
%doi:10.1088/0264-9381/26/23/235003
[arXiv:0810.5163 [gr-qc]].
%25 citations counted in INSPIRE as of 04 Mar 2023
%Copy to ClipboardDownload




%\cite{Bak:2022cnz}
\bibitem{Bak:2022cnz}
D.~Bak, C.~Kim and S.~H.~Yi,
``Structure of deformations in Jackiw-Teitelboim black holes with matter,''
[arXiv:2209.01394 [hep-th]].
%0 citations counted in INSPIRE as of 13 Feb 2023


%\cite{Spradlin:1999bn}
\bibitem{Spradlin:1999bn}
M.~Spradlin and A.~Strominger,
``Vacuum states for AdS(2) black holes,''
JHEP \textbf{11}, 021 (1999)
%doi:10.1088/1126-6708/1999/11/021
[arXiv:hep-th/9904143 [hep-th]].
%157 citations counted in INSPIRE as of 13 Feb 2023



%\cite{Bak:2018txn}
\bibitem{Bak:2018txn}
D.~Bak, C.~Kim and S.~H.~Yi,
``Bulk view of teleportation and traversable wormholes,''
JHEP \textbf{08}, 140 (2018)
%doi:10.1007/JHEP08(2018)140
[arXiv:1805.12349 [hep-th]].
%30 citations counted in INSPIRE as of 11 Aug 2022


%\cite{Harlow:2014yka}
\bibitem{Harlow:2014yka}
D.~Harlow,
``Jerusalem Lectures on Black Holes and Quantum Information,''
Rev. Mod. Phys. \textbf{88}, 015002 (2016)
%doi:10.1103/RevModPhys.88.015002
[arXiv:1409.1231 [hep-th]].
%350 citations counted in INSPIRE as of 14 Feb 2023

%\cite{Crnkovic:1986ex}
\bibitem{Crnkovic:1986ex}
C.~Crnkovic and E.~Witten,
``COVARIANT DESCRIPTION OF CANONICAL FORMALISM IN GEOMETRICAL THEORIES,''
Print-86-1309 (PRINCETON).
%16 citations counted in INSPIRE as of 23 Feb 2023

%\cite{Wald:1995yp}
\bibitem{Wald:1995yp}
R.~M.~Wald,
``Quantum Field Theory in Curved Space-Time and Black Hole Thermodynamics,''
University of Chicago Press,
ISBN 978-0-226-87027-4.
%71 citations counted in INSPIRE as of 23 Feb 2023




%\cite{Cotler:2016fpe}
\bibitem{Cotler:2016fpe}
J.~S.~Cotler, G.~Gur-Ari, M.~Hanada, J.~Polchinski, P.~Saad, S.~H.~Shenker, D.~Stanford, A.~Streicher and M.~Tezuka,
``Black Holes and Random Matrices,''
JHEP \textbf{05}, 118 (2017)
[erratum: JHEP \textbf{09}, 002 (2018)]
%doi:10.1007/JHEP05(2017)118
[arXiv:1611.04650 [hep-th]].
%570 citations counted in INSPIRE as of 04 Mar 2023



 



%\cite{Mandal:2017thl}
\bibitem{Mandal:2017thl}
G.~Mandal, P.~Nayak and S.~R.~Wadia,
%``Coadjoint orbit action of Virasoro group and two-dimensional quantum gravity dual to SYK/tensor models,''
JHEP \textbf{11}, 046 (2017)
%doi:10.1007/JHEP11(2017)046
[arXiv:1702.04266 [hep-th]].
%107 citations counted in INSPIRE as of 04 Mar 2023


%\cite{Mertens:2017mtv}
\bibitem{Mertens:2017mtv}
T.~G.~Mertens, G.~J.~Turiaci and H.~L.~Verlinde,
``Solving the Schwarzian via the Conformal Bootstrap,''
JHEP \textbf{08} (2017), 136
%doi:10.1007/JHEP08(2017)136
[arXiv:1705.08408 [hep-th]].
%240 citations counted in INSPIRE as of 26 Feb 2023



%\cite{Mertens:2018fds}
\bibitem{Mertens:2018fds}
T.~G.~Mertens,
``The Schwarzian theory \textemdash{} origins,''
JHEP \textbf{05}, 036 (2018)
%doi:10.1007/JHEP05(2018)036
[arXiv:1801.09605 [hep-th]].
%116 citations counted in INSPIRE as of 04 Mar 2023


%\cite{Blommaert:2018oro}
\bibitem{Blommaert:2018oro}
A.~Blommaert, T.~G.~Mertens and H.~Verschelde,
``The Schwarzian Theory - A Wilson Line Perspective,''
JHEP \textbf{12} (2018), 022
%doi:10.1007/JHEP12(2018)022
[arXiv:1806.07765 [hep-th]].
%84 citations counted in INSPIRE as of 26 Feb 2023




%\cite{Bak:2007qw}
\bibitem{Bak:2007qw}
D.~Bak, M.~Gutperle and A.~Karch,
``Time dependent black holes and thermal equilibration,''
JHEP \textbf{12}, 034 (2007)
%doi:10.1088/1126-6708/2007/12/034
[arXiv:0708.3691 [hep-th]].
%34 citations counted in INSPIRE as of 04 Mar 2023


%\cite{Kitaev:2018wpr}
\bibitem{Kitaev:2018wpr}
A.~Kitaev and S.~J.~Suh,
``Statistical mechanics of a two-dimensional black hole,''
JHEP \textbf{05} (2019), 198
%doi:10.1007/JHEP05(2019)198
[arXiv:1808.07032 [hep-th]].
%115 citations counted in INSPIRE as of 26 Feb 2023

%\cite{Yang:2018gdb}
\bibitem{Yang:2018gdb}
Z.~Yang,
``The Quantum Gravity Dynamics of Near Extremal Black Holes,''
JHEP \textbf{05} (2019), 205
%doi:10.1007/JHEP05(2019)205
[arXiv:1809.08647 [hep-th]].
%124 citations counted in INSPIRE as of 26 Feb 2023

%\cite{Iliesiu:2019xuh}
\bibitem{Iliesiu:2019xuh}
L.~V.~Iliesiu, S.~S.~Pufu, H.~Verlinde and Y.~Wang,
``An exact quantization of Jackiw-Teitelboim gravity,''
JHEP \textbf{11} (2019), 091
%doi:10.1007/JHEP11(2019)091
[arXiv:1905.02726 [hep-th]].
%72 citations counted in INSPIRE as of 26 Feb 2023

%\cite{Lin:2022rbf}
\bibitem{Lin:2022rbf}
H.~W.~Lin,
``The bulk Hilbert space of double scaled SYK,''
JHEP \textbf{11}, 060 (2022)
%doi:10.1007/JHEP11(2022)060
[arXiv:2208.07032 [hep-th]].
%15 citations counted in INSPIRE as of 04 Mar 2023


%\cite{Kitaev:2017hnr}
\bibitem{Kitaev:2017hnr}
A.~Kitaev,
``Notes on $\widetilde{\mathrm{SL}}(2,\mathbb{R})$ representations,''
[arXiv:1711.08169 [hep-th]].
%72 citations counted in INSPIRE as of 04 Mar 2023
 


%%\cite{Almheiri:2020cfm}
%\bibitem{Almheiri:2020cfm}
%A.~Almheiri, T.~Hartman, J.~Maldacena, E.~Shaghoulian and A.~Tajdini,
%``The entropy of Hawking radiation,''
%Rev. Mod. Phys. \textbf{93}, no.3, 035002 (2021)
%%doi:10.1103/RevModPhys.93.035002
%[arXiv:2006.06872 [hep-th]].
%%293 citations counted in INSPIRE as of 16 Aug 2022
%

%%\cite{Maldacena:2001kr}
%\bibitem{Maldacena:2001kr}
%J.~M.~Maldacena,
%``Eternal black holes in anti-de Sitter,''
%JHEP \textbf{04}, 021 (2003)
%%doi:10.1088/1126-6708/2003/04/021
%[arXiv:hep-th/0106112 [hep-th]].
%%1181 citations counted in INSPIRE as of 15 Aug 2022


%%\cite{Breitenlohner:1982jf}
%\bibitem{Breitenlohner:1982jf}
%P.~Breitenlohner and D.~Z.~Freedman,
%``Stability in Gauged Extended Supergravity,''
%Annals Phys. \textbf{144}, 249 (1982).
%%doi:10.1016/0003-4916(82)90116-6
%%1593 citations counted in INSPIRE as of 11 Aug 2022

%%\cite{Witten:2001ua}
%\bibitem{Witten:2001ua}
%E.~Witten,
%``Multitrace operators, boundary conditions, and AdS / CFT correspondence,''
%[arXiv:hep-th/0112258 [hep-th]].
%%545 citations counted in INSPIRE as of 13 Aug 2022
%

%%\cite{Goto:2017olq}
%\bibitem{Goto:2017olq}
%K.~Goto and T.~Takayanagi,
%``CFT descriptions of bulk local states in the AdS black holes,''
%JHEP \textbf{10}, 153 (2017)
%%doi:10.1007/JHEP10(2017)153
%[arXiv:1704.00053 [hep-th]].
%%35 citations counted in INSPIRE as of 24 Aug 2022




%%\cite{Brown:2019rox}
%\bibitem{Brown:2019rox}
%A.~R.~Brown, H.~Gharibyan, G.~Penington and L.~Susskind,
%``The Python\textquoteright{}s Lunch: geometric obstructions to decoding Hawking radiation,''
%JHEP \textbf{08}, 121 (2020)
%%doi:10.1007/JHEP08(2020)121
%[arXiv:1912.00228 [hep-th]].
%%51 citations counted in INSPIRE as of 26 Aug 2022

%%\cite{Engelhardt:2021mue}
%\bibitem{Engelhardt:2021mue}
%N.~Engelhardt, G.~Penington and A.~Shahbazi-Moghaddam,
%``A world without pythons would be so simple,''
%Class. Quant. Grav. \textbf{38}, no.23, 234001 (2021)
%%doi:10.1088/1361-6382/ac2de5
%[arXiv:2102.07774 [hep-th]].
%%15 citations counted in INSPIRE as of 26 Aug 2022







%%\cite{Maldacena:2017axo}
%\bibitem{Maldacena:2017axo}
%J.~Maldacena, D.~Stanford and Z.~Yang,
%``Diving into traversable wormholes,''
%Fortsch. Phys. \textbf{65}, no.5, 1700034 (2017)
%%doi:10.1002/prop.201700034
%[arXiv:1704.05333 [hep-th]].
%%270 citations counted in INSPIRE as of 18 Aug 2022
%
%
%
%%\cite{Skenderis:2002wp}
%\bibitem{Skenderis:2002wp}
%K.~Skenderis,
%``Lecture notes on holographic renormalization,''
%Class. Quant. Grav. \textbf{19}, 5849-5876 (2002)
%%doi:10.1088/0264-9381/19/22/306
%[arXiv:hep-th/0209067 [hep-th]].
%%1181 citations counted in INSPIRE as of 17 Aug 2022
%
%
%%\cite{Takahasi:1974zn}
%\bibitem{Takahasi:1974zn} 
%  Y.~Takahasi and H.~Umezawa,
%  ``Thermo field dynamics,''
%  Collect.\ Phenom.\  {\bf 2}, 55 (1975).
%  %%CITATION = CLPNA,2,55;%%
%  %418 citations counted in INSPIRE as of 24 May 2018
%



%%\cite{Bak:2002rq}
%\bibitem{Bak:2002rq}
%D.~Bak,
%``SUPERSYMMETRIC BRANES IN THE MATRIX MODEL OF A PP WAVE BACKGROUND,''
%Phys. Rev. D \textbf{67}, 045017 (2003)
%%doi:10.1103/PhysRevD.67.045017
%[arXiv:hep-th/0204033 [hep-th]].
%%95 citations counted in INSPIRE as of 11 Aug 2022


 



\end{thebibliography}
\end{document}